\documentstyle[aps,preprint]{revtex}
\draft \tighten
\input epsf
\begin{document}
\draft
\newcommand{\picdir}[1]{./#1}     
%
%
\def\pbg{\hbox{${\mathbf \pi\beta\gamma}$}}
\def\tsigma{\tilde{\sigma}}
\def\lesssim{\mathrel{\hbox{\rlap{\hbox{\lower4pt\hbox{$\sim$}}}\hbox{$<
$}}}}
\def\gtrsim{\mathrel{\hbox{\rlap{\hbox{\lower4pt\hbox{$\sim$}}}\hbox{$>$
}}}}
\preprint{CITA-2001-12, SU-ITP-01/26,  hep-th/0106179}
\date{June 19, 2001}
\title{\Large\bf  Tachyonic Instability and Dynamics of Spontaneous Symmetry
Breaking}
\author{Gary Felder$^{1}$,   ~Lev Kofman$^{2}$, and ~Andrei Linde$^{1}$  }
\address{${}^1$Department of Physics, Stanford
University, Stanford, CA 94305, USA} \address{${}^2$ CITA, University of
Toronto, 60 St. George Street, Toronto, ON M5S 1A7, Canada}
  \maketitle
\begin{abstract}
Spontaneous  symmetry breaking usually occurs due to the tachyonic
(spinodal) instability of a scalar field near the top of its effective
potential at $\phi = 0$. Naively, one might expect the field $\phi$ to fall
from the top of the effective potential and then experience a long stage of
oscillations with  amplitude $O(v)$ near the minimum of the effective
potential at $\phi = v$ until it gives its energy to particles produced
during these oscillations. However, it was recently found that the tachyonic
instability rapidly converts most of the potential energy $V(0)$ into the
energy of colliding classical waves of the scalar field. This conversion,
which was called ``tachyonic preheating,'' is so efficient that symmetry
breaking typically completes within a single oscillation of the field
distribution as it rolls towards the minimum of its effective potential
\cite{GBFKLT}.   In this paper we give a detailed description of tachyonic
preheating and show that the dynamics of this process crucially depend on
the shape of the effective potential near its maximum. In the simplest
models where $V(\phi) \sim - m^2\phi^2/2$ near the maximum, the process
occurs solely due to the tachyonic instability, whereas in the theories
$-\lambda\phi^n$ with $n > 2$ one encounters a combination of the effects of
tunneling, tachyonic instability and bubble wall collisions.
\end{abstract}

\newpage

\section{Introduction}

Since the beginning of the 1970's, spontaneous symmetry breaking (SSB)
has been  a basic feature of all realistic theories of elementary particles. It is
discussed in every book on quantum field theory, so one might expect
the theory of this effect to be  well understood. However, until
very recently this was not the case.

The standard picture of SSB that many people have in mind looks as follows.
The field initially stays at the top of the effective potential $V(\phi)$
at $\phi = 0$ like a ball at the top of a hill. Then some small external
force pushes it to the right or the left. Even if this force is
infinitesimally small, it is enough for the field to start falling down in
the direction in which it was pushed. The field then oscillates near the minimum
of its effective potential $V(\phi)$ at $|\phi| = v$, and eventually the whole
universe becomes filled by a homogeneous scalar field $|\phi| = v$.  As an
example of this process one can imagine a piece of ferromagnetic
material being magnetized under the influence of a very small
external magnetic field.

That is why many authors who studied SSB assumed that the field
$\phi$ was initially slightly displaced from the maximum of $V(\phi)$. Then
they studied classical rolling of the field $\phi$ from this displaced
state and the growth of quantum fluctuations on top of the homogeneous
classical field. It was generally thought that the stage of
oscillations of a homogeneous classical field $\phi$ with amplitude
$|\Delta \phi| \sim v$ would last for a very long time until it produced
elementary particles that drained the energy of the  classical oscillations.
One can talk about spontaneous symmetry breaking only when the field $\phi$
settles down at $|\phi| \sim v$ and starts oscillating near this state with
an amplitude much smaller than $v$: $|\Delta \phi| \ll v$. The most
efficient process that was previously known to convert the energy of a  homogeneously  oscillating scalar field into the energy of elementary particles and make the amplitude of the oscillations small was parametric resonance \cite{KLS}, but  in most cases studied in the literature the homogeneous component of the field makes several dozen  oscillations before the process completes.

However, in a recent paper by Felder, Garc\'\i a-Bellido,  Greene,
Kofman,  Linde and  Tkachev \cite{GBFKLT} it was shown that typically
spontaneous symmetry breaking completes within a single oscillation of the
scalar field.  One key observation made in \cite{GBFKLT} was that nobody
pushes the field $\phi$ from the top of the effective potential in the
early universe, so the usual picture of a homogeneously oscillating scalar
field is incorrect in application to SSB.  In those cases when the initial value of the homogeneous component of the field $\phi$ is  close  to zero,  quantum fluctuations rather than the classical rolling of the homogeneous field $\phi$ dominate the dynamics.  

Usually SSB occurs  because of
the presence of tachyonic mass terms such as $-m^2 \phi^2/2$ in the
effective
potential, so that $V'' = -m^2 < 0$. Long wavelength quantum fluctuations
$\phi_k$ of the field $\phi$ with momenta $k < m$ grow
exponentially, $\phi_k \sim \exp (t\sqrt{m^2- k^2})$. When these
fluctuations become large they can be interpreted as classical waves of
the scalar field. Spontaneous symmetry breaking  occurs  when the total
amplitude of these fluctuations grows up to $v$. Because all modes with $k
< m$ are growing, SSB occurs {\it locally} on a scale somewhat greater
than $m^{-1}$. Later on, this scale gradually increases. Inhomogeneities of
the scalar field absorb some part of the energy $V(0)$, which  suppresses
the amplitude of the scalar field oscillations. As a result,  the field
$\phi$ that appears after SSB is relatively homogeneous on the scale somewhat greater than $m^{-1}$, and the amplitude of its
oscillations $\Delta \phi(x,t)$ about the state $|\phi| \sim v$ is
substantially smaller than $v$. Thus, contrary to naive expectations, a
prolonged stage of oscillations of a homogeneous component of the scalar
fields during SSB  usually does not exist.

 As we will show in this paper, one reaches a similar conclusion 
even if the field $\phi$ initially has been slightly displaced from the top of the
potential (no SSB). With this initial condition, the homogeneous
background field  decays within few oscillations  due to the
broad parametric resonance enhanced by the tachyonic regime.

The process of rapid transfer
of the energy of the scalar field $V(0)$ into the energy of its
inhomogeneous oscillations due to tachyonic instability was called {\it tachyonic preheating} \cite{GBFKLT}. One should distinguish between the tachyonic preheating and spinodal (tachyonic) instability, which occurs at the very beginning of this process. The first stages of the process of SSB  related to tachyonic (spinodal) instability can be studied by relatively simple methods. However,
very soon the process becomes exceedingly complicated.  When the field grows sufficiently large, one should take into account nonlinear effects.
Oscillations of the field  can trigger an explosive process of particle production due to parametric resonance \cite{KLS}, which can be especially efficient in our case because of the  tachyonic instability. Particles (waves) of the classical field produced in this process begin interacting with each other (rescattering). At this stage even advanced methods based on the Hartree \cite{KLS} or
$1/N$ \cite{Boyan} approximations fail to describe the situation correctly.
In addition,  from the very beginning of the process there may be
production of topological defects, which cannot be described by
perturbation theory. One might expect that since this is a nonperturbative
phenomenon it cannot materially affect the process of SSB. As we will see, however, the production of topological defects is not a small  correction but an important feature of SSB. Topological defects, like other inhomogeneities generated by
tachyonic instability,  drain  the energy of the scalar field rolling down
to the minimum of the effective potential. By doing so, they diminish  the
amplitude of subsequent oscillations of the scalar field.

 There is an extensive literature describing SSB, spinodal instability and the production of topological defects during  high temperature phase transitions in cosmology \cite{KL72,KL76}. To study these issues one should find how the temperature changes in the early universe \cite{book} and
use numerical methods to find out how symmetry
breaking occurs in an expanding universe with a time-dependent temperature. Many
interesting results have been obtained in this direction, see e.g.
\cite{VilShell}. However, most of these results were strongly
model-dependent because the answers crucially depend on the ratios between
masses of the particles, their coupling constants, the temperature of the
universe and the rate of expansion. To avoid this problem, in this
paper we will concentrate on the simplest possibility when the temperature
was zero from the very beginning and the field was standing on the top of
the effective potential. This will allow us to study basic features of the
process of spontaneous symmetry breaking in its pure form without extra
complications related to high temperature effects and cosmological
evolution.\footnote{A discussion of SSB and tachyonic preheating in the early universe will be contained in \cite{FKL2,GBFKL}.}

Even in this regime, the theory of SSB remains extremely complicated since
for its investigation one should go
beyond perturbation theory. Fortunately, during the last few years
new methods of lattice simulations have been developed. They are
based on the observation that quantum states of bose fields with
large occupation numbers can be interpreted as classical waves and
their dynamics can be fully analysed by solving relativistic wave
equations on a lattice \cite{lattice,FT}.  Similar methods were used in
\cite{Grig,Bjorken,Wil,Cal,Bettencourt:2000jv} in application to sphaleron
effects, the formation of disoriented chiral condensates, and the problem of
topological defect production in the early universe.  In our paper, which
extends the previous work \cite{GBFKLT}, we will
further develop these methods.   

Usually the main output of lattice simulations is the calculation of
correlation functions, Wilson loops, etc. A significant advantage of our
methods  is that the
semi-classical nature of the effects under investigation allows us
to have a clear visual picture of all the processes involved. One can
really {\it see} the process of spontaneous symmetry breaking
\cite{GBFKLT}, which helps enormously in understanding the nature of
this effect.   That
is why this paper is accompanied by many figures that show the development
of symmetry breaking in various
models.

In addition to the simplest models with $V''(0) = -m^2 < 0$, we will study some  models where the curvature of the effective potential near $\phi = 0$ is negative, but it vanishes at $\phi = 0$. This happens in such  theories as
$-\lambda\phi^4$ or $-\lambda  \phi^3$. (Potentials of the type of $-\lambda\phi^3$ appear in the simplest SUSY motivated models of hybrid inflation.)
 As we will see, the development of
tachyonic instability in such models is accompanied by bubble formation and
growth and bubble wall collisions.   Rather interestingly, in these theories bubble formation occurs
via tunneling even though there is no potential barrier in $V(\phi)$
\cite{Linde1,LW}. Moreover in the theory $-\lambda\phi^3$ this process occurs even
though there are no instantons describing bubble formation in this theory.
To understand this process one should use the stochastic approach to tunneling
developed in \cite{Linde2}.

Section \ref{theory} describes the basic theory of spontaneous symmetry
breaking and tachyonic instability, focusing particularly on the simplest
example of a negative quadratic potential. In this section we also discuss
the definition of occupation number used throughout the paper to describe
the growth of fluctuations. Section \ref{displacement} generalizes the
theory to a broader class of potentials and to the case where the
homogeneous field begins displaced from the maximum of the
potential. Section \ref{quadratic} presents the results of our numerical simulations for the simplest SSB model, a single field with a quadratic tachyonic term (i.e. $V \sim -  m^2 \phi^2/2$). Section \ref{perturbative} compares our results with results obtained from perturbative calculations and shows how and when the perturbative calculations break down. Section \ref{complexsection} extends our numerical calculations to the case of a complex field with a quadratic tachyonic term.

The next two sections discuss the somewhat more complicated situation that
arises when the tachyonic mass is $\phi$ dependent and vanishes at
$\phi=0$. Section \ref{quarticsection} discusses quartic potentials ($V
 \sim -  \lambda \phi^4/4$) and explains how SSB occurs through tunneling and bubble formation in such models. Section \ref{cubic} discusses cubic potentials where the behavior is in some ways intermediate between that of the quadratic and quartic cases.

In the concluding section we summarize our results and briefly discuss their application to various cosmological scenarios, including hybrid inflation, new inflation, brane inflation, and the recently proposed  ekpyrotic/pyrotechnic
 universe scenarios. Finally there is an appendix that provides details on our lattice calculations and lists the parameters used for each of the simulations described in the paper.

\section{Tachyonic Instability  and spontaneous symmetry breaking}\label{theory}

The simplest model of spontaneous symmetry breaking is based on the
theory with the effective potential
\begin{equation}\label{aB1}
V (\phi) = {\lambda\over 4}(\phi^2-{  v}^2)^2 \equiv   - {m^2\over
2}\phi^2 + {\lambda\over 4} \phi^4 + {m^4\over
4\lambda}\ ,
\end{equation}
where  $m^2 \equiv \lambda v^2$ and $\lambda \ll 1$.  $V(\phi)$ has
a maximum at $\phi = 0$ with curvature  $V'' \equiv V_{\phi\phi} = -m^2$
 and a minimum at $\phi = \pm v$.

The development of tachyonic instability in this model depends on
the initial conditions. We will assume that initially the symmetry
is completely restored so that the field $\phi$ does not have any
homogeneous component, i.e. $\langle \phi \rangle= 0$. But then
$\langle \phi \rangle$ remains zero at all later stages and for
the investigation of SSB one needs to
find the spatial distribution of the field $\phi(x,t)$.  To
avoid this complication, many authors assume that there is a small
but finite initial homogeneous background field $\phi(t)$, and
even smaller quantum fluctuations $\delta\phi(x,t)$ that grow on
top of it. This approximation may provide some interesting
information, but quite often it is inadequate. In particular, it
does not describe the creation of topological defects, which, as
we will see, is not a small nonperturbative correction but an
important part of the problem.

Let us consider equation for the scalar field fluctuations in the model (\ref{aB1}):
\begin{equation}
\ddot \phi_k +(k^2 + V'')\phi_k=0 \ .
\end{equation}  
For definiteness, we suppose that  the mode functions describing quantum fluctuations in the symmetric phase $\phi=0$ at the moment close to  $t = 0$ are the same as for a massless field,
$\phi_k ={1 \over \sqrt{2 k}}e^{-ikt +i{\vec k \vec
x}}$. Then at $t = 0$ we `turn on' the term $-m^2\phi^2/2$
corresponding to the negative mass squared $-m^2$. The modes with
$k = |{\vec k}|  < m$ grow  exponentially. 
Initial dispersion of all growing fluctuations with $k < m$ was given by \begin{equation}
\langle \delta\phi^2 \rangle
 =  \int\limits_0^{m}  {  dk^2  \over 8\pi^2 } = {m^2\over 8 \pi^2} \ ,
\label{aBBB}
\end{equation}
and the average initial amplitude of all fluctuations with $k < m$ was given by 
\begin{equation}
 \delta\phi
 =  {m \over 2 \pi } \ .
\label{aBBBB}
\end{equation}

The  dispersion of the growing modes at  $t > 0$ is given by
\begin{equation}
\langle \delta\phi^2 \rangle
 =  \int\limits_0^{m}  {  dk^2  \over 8\pi^2 }\, e^{2t\sqrt{m^2- k^2}}
 = {e^{2mt}(2mt-1)+1\over 16\pi^2 t^2}  \ .
\label{aBB}
\end{equation}
This means that the average amplitude $\delta\phi(k)$ of quantum fluctuations with momenta   $\sim k$ initially was $\delta\phi(k) \sim k/2\pi$, and then it started growing as $e^{t\sqrt{m^2- k^2}}$.

To get a qualitative understanding of the process of spontaneous
symmetry breaking, instead of many growing waves with momenta $k <
m$ in (\ref{aBB}) let us consider first a single sinusoidal wave
$\delta\phi = \Delta(t) \cos kx$ with $k \sim m$ and with initial
amplitude $\Delta(t) \sim {m\over 2\pi}$ in one-dimensional space (so that the average value of $(\delta\phi)^2$ corresponds to ${m^2\over 8 \pi^2}$). The
amplitude of this wave grows exponentially until it becomes
${\cal O}(v) \sim m/\sqrt \lambda$.  This leads to the division
of the universe into domains of size ${\cal O}(m^{-1})$ in which
the field changes from ${\cal O}(v)$ to ${\cal O}(-v)$. The
gradient energy density of domain walls separating areas with
positive and negative $\phi$ will be $\sim k^2\delta\phi^2=
O(m^4/\lambda)$. This energy is of the same order as the total
initial potential energy of the field $V(0) = m^4/4\lambda$. This
is one of the reasons why any approximation based on perturbation
theory and ignoring topological defect production cannot give a
correct description of the process of spontaneous symmetry
breaking.

Thus a substantial part of the energy $V(0)$ is
transferred to the gradient energy of the field $\phi$ when it
rolls down to the minimum of $V(\phi)$. Because the initial state
contains many quantum fluctuations with different phases growing
at different rates, the resulting field distribution is a gaussian random field with varying spectrum. It cannot coherently give all of its gradient energy back and
return to its initial state $\phi = 0$.  This is one of the
reasons why spontaneous symmetry breaking and the main stage of
preheating in this model may occur within a single oscillation of
the field~$\phi$.

Meanwhile if one were to make the usual assumption that initially
there exists a small homogeneous background field $\phi \ll v$
with an amplitude greater than the amplitude of the growing
quantum fluctuations $\delta\phi$, so that $m/2\pi \ll \phi <
m/\sqrt \lambda$, one would find out that when $\phi$ falls to the
minimum of the effective potential the gradient energy of the
fluctuations remains relatively small. In some situations this could lead one to falsely conclude that the field will experience many
fluctuations before it relaxes near the minimum of $V(\phi)$. To
avoid this error, we need to perform a complete study of the
growth of all tachyonic modes and their subsequent interaction
without making this simplifying assumption about the existence of
a homogeneous field $\phi$.

The tachyonic growth of all fluctuations with $k <
m$ continues until $\sqrt{\langle
\delta\phi^2 \rangle}$ reaches the value $\sim v/2$, since at
$\phi \sim v/\sqrt 3$ the curvature of the effective potential
vanishes and instead of tachyonic growth one has the usual
oscillations of all the modes. Eq. (\ref{aBB}) shows that this happens
within a time $t_*
\sim {1\over 2 m} \ln{C\over \lambda}$, where $C \sim 10^2$.

A convenient tool for studying this process   is the spectrum of  the growing
quantum fluctuations.  Rather than using the usual power spectrum $\vert\phi_k\vert^2$, however, we find it more informative to
investigate the  occupation number $n_k$
of produced particles  \cite{KLS}.  Indeed, in situations where the
number of particles is well defined (and this always happens at the end of
the process) the occupation number $n_k$ is an adiabatic invariant, i.e. it
does not change during the field oscillations unless some dramatic changes
occur to the system. The standard definition of the occupation number which
was extensively used in the theory of preheating \cite{KLS}, and which is
valid for $m^2 \geq 0$, is
\begin{equation}\label{number}
n_k={\omega_k\over 2} \left( { |\dot \phi_k|^2 \over \omega_k^2}
+|\phi_k|^2 \right)-{1\over 2} \ .
\end{equation}
However, this definition does not work in the tachyonic regime when the
effective mass squared of the field $\phi$ becomes negative since then
$\omega_k = \sqrt{k^2 + m^2}$ becomes imaginary. Strictly speaking, $n_k$ should not be interpreted as  the occupation number of particles during the tachyonic
regime. One may still formally calculate the function $n_k$ in this regime using  either $\sqrt{k^2 +|m^2|}$ or   $|k|$ instead of $\omega_k= \sqrt{k^2 + m^2}$ in the expression for $n_k$   whenever $m^2<0$. The choice between $\sqrt{k^2 +|m^2|}$ and $|k|$ is arbitrary, and it does not change any of the final physical results.   The spectra shown in this paper used $\omega_k=\vert k\vert$ in the tachyonic regime, with the exception of Fig. 1 and Fig. 2, where we used $\omega_k= \sqrt{k^2 + |m^2|}$.   The formally defined quantity $n_k$ can be interpreted as the
occupation number of particles after  the end of the tachyonic regime, when
$m^2 \geq 0$. Moreover, for all nonvanishing momenta these two quantities
 match very well when one switches from the tachyonic regime to the normal
one. Indeed, whereas the value of $\omega_k$ changes
during the process, the exponential growth of $n_k$ is mainly determined by $|\dot
\phi_k|$ and $|\phi_k|$, which  do not change
strongly during the switch between the tachyonic regime and the normal one.
Therefore the function $n_k$ is very convenient and informative during the
whole process. When one calculates $n_k$ during the tachyonic regime, one can
get a  good idea of the number of particles that will emerge at the
end of this regime where the usual particle interpretation becomes
possible. In this sense we will interpret the function $n_k$ defined above
as the occupation number of particles  in both regimes.

An additional caveat of this interpretation is that when one calculates $\phi_k$, one does not distinguish between the contribution to this quantity from small perturbations and from topological defects. As a result, in the presence of topological defects one can somewhat overestimate the number of produced particles. The error, however, is not very large, especially in theories where instead of domain walls we have strings or monopoles. Moreover, eventually topological defects disappear and release their energy in the form of produced particles, and the standard interpretation of $n_k$ becomes completely valid.

The exponential growth
of fluctuations during the tachyonic regime  can be interpreted as the
growth
of the occupation number of particles with $k \ll m$. Using the estimates
given above, one can show that $n_k$ for $k \ll m$
at the time $t_*$ grows up to
\begin{equation}\label{occlam}
n_k \sim \exp(2mt_*)   =  {\cal O}(10^2)\, \lambda^{-1} \gg 1\,.
\end{equation}
The time $t_*
\sim {1\over 2 m} \ln{C\over \lambda}$ depends only logarithmically not
only on $\lambda$, but, more generally, on the choice of the initial
distribution of quantum fluctuations. As we see, for small $\lambda$ the
fluctuations with $k \ll m$ acquire very large occupation numbers.  More importantly, such fluctuations will have large amplitude  and  will be in squeezed state. The general solution for these fluctuations will contain two terms, $A e^{\omega t}$ and $B e^{-\omega t}$, but after a
 short time only the growing mode $A e^{\omega t}$ survives. Therefore,  independently of the initial phases of quantum fluctuations, the only modes with $k < m$ that survive  after the  beginning of the tachyonic regime will coherently grow, and their amplitude  will become extremely large. That is why these modes can be interpreted as classical waves and can be studied by computer simulations using the methods of \cite{lattice,FT}.

The dominant contribution to $\langle \delta\phi^2 \rangle$ in Eq.
(\ref{aBB}) at the moment $t_*$ is given by the modes with wavelength $l_*
\sim 2\pi
k_*^{-1} \sim \sqrt 2 \pi m^{-1} \ln^{1/2} {(C/\lambda)} >
m^{-1}$, where $C = {\cal O}(10^2)$. As a result, at the moment when the
fluctuations of the field $\phi$ reach the minimum of the
effective potential, $\sqrt{\langle\delta\phi^2\rangle} \sim v$, the
field distribution looks rather homogeneous on a scale $l \lesssim
l_*$. On average, one still has $\langle \phi\rangle = 0$. This
implies that the universe becomes divided into domains with two
different types of spontaneous symmetry breaking, $\phi \sim \pm
v$. The typical size of each domain is $l_*/2 \sim {\pi\over \sqrt
2 } \ m^{-1} \ln^{1/2}{C\over \lambda}$, which is slightly greater than $m^{-1}$. At later stages the domains grow in size and percolate (eat each other),
and SSB becomes established on a macroscopic scale.

Of course, these are just simple estimates that should be
followed by a detailed quantitative investigation. When the field
rolls down to the minimum of its effective potential its
fluctuations scatter off each other as classical waves due to the $\lambda \phi^4$ interaction. It is
difficult to study this process analytically, but fortunately one
can do it numerically using the method of lattice simulations
developed in \cite{lattice,FT}.

Before describing the results of our lattice simulations, we would like to discuss the setting of the problem, the choice of the initial conditions and some other aspects of tachyonic instability in a more general class of theories, including the theories with $V(\phi) \sim -\lambda\phi^n$.

\section{ Tachyonic instability in a more general class of theories and the role of initial displacement}\label{displacement}

As we already emphasized, spontaneous symmetry breaking usually occurs due to quantum fluctuations  when the field $\phi$ falls from an exactly symmetric state $\phi = 0$. However, it is still very instructive to find out what happens when the field falls down from some state with $\phi_0 \not = 0$. By doing this, we will get an additional insight in the nature of tachyonic instability. We will also be able to compare our results with the results of earlier works on spontaneous symmetry breaking.

 Consider the behavior of the fluctuations $\delta \phi$ with momentum $k \ll m$. An important observation is that these fluctuations satisfy the same equation of motion as $\dot \phi$:
\begin{equation}\label{deltaphi}
\ddot{ \phi_k}   =  -V''(\phi)\,  \phi_k \,.
\end{equation}
A general solution of this equation for $V = -m^2\phi^2/2$ is $ \phi_k(t) = a_1 e^{mt} + a_2 e^{-mt}$. Similarly, for $\dot\phi$ one has $\dot\phi = b_1 e^{mt} + b_2 e^{-mt}$. At  $t \gg m^{-1}$ only the growing mode survives, and the ratio  
$\phi_k/\dot\phi$ becomes constant. This rule holds for other types of tachyonic potentials as well. Thus one can investigate the amplification of the long wavelength perturbations of the scalar field $\phi$ in a very easy way. Instead of solving equations for $ \phi_k$ in a time-dependent background  $\phi(t)$, one can find how $\dot\phi(t)$ changes in time. We will use this trick here and in the next section.

Consider the theory
\begin{equation}
V(\phi) = V_0- \lambda\phi^n/2  \ .
\end{equation}
Suppose the field $\phi$ begins rolling down from $\phi_0$. Energy conservation implies that
\begin{equation}
\dot \phi^2/2- \dot \phi_0^2/2  = V(\phi_0) - V(\phi) \ .
\end{equation}
We will assume for simplicity that in the beginning, at $\phi = \phi_0$, the field moves with the same velocity as if it were falling  with vanishing total energy from $\phi=0$ (this assumption does not make any difference for motion at
$\phi \gg \phi_0$). Then one has
\begin{equation}
\dot \phi^2/2   =   - V(\phi)=  \lambda\phi^n/2 \ .
\end{equation}
Thus
\begin{equation}\label{rrr}
\dot \phi   = \sqrt \lambda \phi^{n/2} \ .
\end{equation}
The solution is
\begin{equation}
 \phi   = \left(\phi_0^{{2-n\over 2}} - \sqrt \lambda t \left({n-2\over 2}\right)\right)^{-{2\over n-2}}.
\end{equation}
The most important result is Eq. (\ref{rrr}). 
In a more general case of nonvanishing total energy we have
\begin{equation}\label{rrrr}
\dot \phi   = \sqrt{\lambda (\phi^n- \phi_0^n) }\ ,
\end{equation}
where $\phi_0$ is an initial field value where $\dot \phi=0$.
It implies that the tachyonic fluctuations with small momenta in the long time limit 
($\phi \ll \phi_0$)
grow as follows:
\begin{equation}\label{rrr2}
 \phi_k =  C\, \phi^{n/2}  \ ,
\end{equation}
where $C$ is some constant.

This means, in particular, that in the theory with the potential $-  \phi^2 $ the long wavelength fluctuations grow just like the field itself, ${ \phi_k\over  \phi} = const$.  Meanwhile for the theory with $-  \phi^3$ the fluctuations grow faster,  ${ \phi_k\over  \phi} \sim \phi^{1/2}$, and for the theory $-  \phi^4$ they grow even faster, ${ \phi_k\over  \phi} \sim \phi$.

Returning to the theory $   - {m^2\over
2}\phi^2 + {\lambda\over 4} \phi^4 + {m^4\over
4\lambda}$, we find that the potential is tachyonic ($V'' < 0$) for $0 < \phi < v/\sqrt 3$, and it can be approximately represented as $   - {m^2\over
2}\phi^2$ for $0 < \phi \lesssim v/2$. When the field $\phi$ grows from $\phi_0 \ll v$ to $v/\sqrt 3$, the speed of the field $\dot \phi$ grows from $m\phi_0$ to   ${\sqrt 5\over 6}mv$. Consequently, the amplitude of density perturbations grows by a factor $\sim  {\sqrt 5\over 6}{v\over \phi_0}$, and the occupation numbers $n_k$ of particles for $k \ll m$ grow by a factor $O\left({v^2\over 5 \phi_0^2}\right)$.

Clearly, one has the largest amplification if one starts as close to $\phi_0 = 0$ as possible. However, if   $\phi_0 \ll {m\over 2\pi}$, where ${m\over 2\pi}$ is the average amplitude of the long wavelength quantum fluctuations with momentum $k < m$ (which grow almost as fast as the homogeneous mode), then the development of $\phi_0$ gives no information on the process of spontaneous symmetry breaking.  In this case instead of $\phi_0$  one would need to study all growing modes with $k < m$, just as in the case $\phi_0 = 0$. This is what we are doing in the main part of this paper.

 Equation for fluctuations in the model $ V(\phi)=  - {m^2\over
2}\phi^2 + {\lambda\over 4} \phi^4 + {m^4\over 4 \lambda}$ is
\begin{equation}\label{fluc}
\ddot  \phi_k + \left( k^2 -m^2 +3 \lambda \phi^2 \right)  \phi_k=0\    ,
\end{equation}
This equation should be solved simultaneously with the equation for the background field $\phi(t)$
\begin{equation}\label{back}
\ddot \phi  -m^2 \phi  + \lambda \phi^3=0\   ,
\end{equation}
Equation (\ref{fluc})  is the Lame equation \cite{GKLS}. Its solutions depend on the  dimensionless parameters $\sqrt{\lambda} \phi_0/m$.
In the context of the chaotic inflationary model, where the field $\phi(t)$
is rolling from its large initial value $\phi \sim M_p$,
this parameter usually was taken to be large \cite{GKLS}.
In the context of the theory of spontaneous symmetry  breaking we are dealing with the opposite case when $\phi_0$ is close to zero.

The description of the growth of perturbations  
in the model  $  - {m^2\over
2}\phi^2 + {\lambda\over 4} \phi^4 + {m^4\over 4\lambda}$ 
is a straightforward generalization of the theory of parametric resonance in the
model $ {\lambda\over 4} \phi^4$, which has been studied using
the stability/instability chart of the Lame equation \cite{GKLS}. 
The presence of the  negative mass term  adds an additional
instability band at $k \lesssim m$. 
The characteristic exponent $\mu$ in this new zone is significantly greater
that in the higher zones because of the tachyonic effect.  
Thus, the tachyonic parametric resonance will be dominant.

 When the field rolls towards the minimum of $V(\phi)$, the occupation numbers 
$n_k$, calculated from the solutions $ \phi_k$ of equation (\ref{fluc}),
become large. However, for $\phi_0 > {m\over 2\pi} $  the field fluctuations  do not grow large enough to dominate the energy density immediately after the rolling to the minimum of the effective potential. To find out what happens in this case, we will describe, as an example, the evolution of the occupation numbers of the modes  $\phi_k$ with different momenta $k$ in the model (\ref{aB1}) with $\lambda = 10^{-4}$ if the field rolls from $\phi_0 = 0.01 v$ (which is larger than ${m\over 2\pi}$ in this model).

\begin{figure}[Fig001]
\centering \leavevmode\epsfysize=7cm
\epsfbox{\picdir{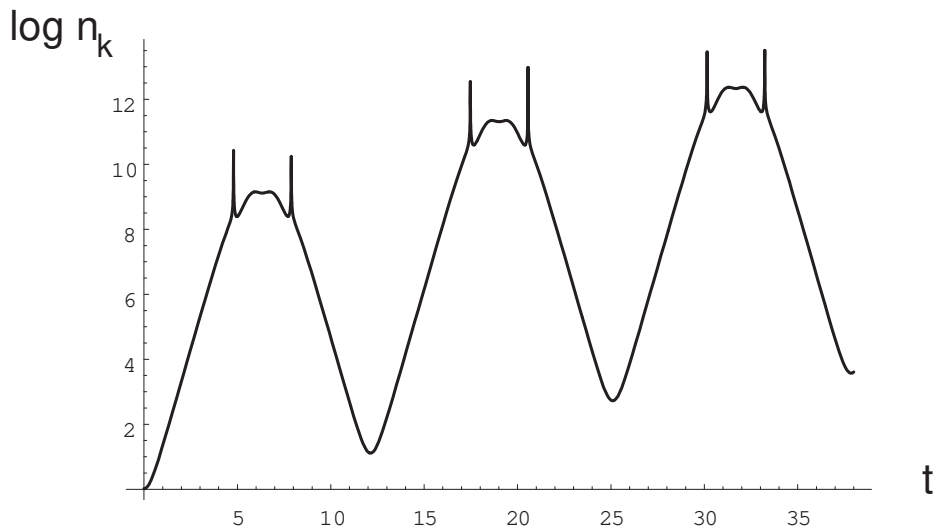}}

\

\caption[Fig001]{\label{k=0} { Evolution of the occupation numbers for the fluctuations with $k \ll m$ in the model $ V(\phi)=  - {m^2\over
2}\phi^2 + {\lambda\over 4} \phi^4 + {m^4\over 4\lambda}$}}
\end{figure}

 Consider first a mode $\phi_k$ with $k \ll m$. In the beginning, when the field $\phi$ rolls from $\phi = \phi_0$ to $\phi = v/\sqrt 3$, this mode grows faster than any other fluctuations, just as we expected, see Fig. \ref{k=0}. During this time interval, the occupation number becomes $\sim e^9$, which is in good agreement with our estimate $O\left({v^2\over 5 \phi_0^2}\right) \sim 2\times 10^3$. Then the field reaches the bottom of the effective potential, goes somewhat beyond this point, bounces back, and again approaches the tachyonic region $\phi < v/\sqrt 3$. Until the field becomes smaller than $v/\sqrt 3$, the occupation number of particles with $k \ll m$ does not change much. But then it {\it decreases} almost to the same value from which we started our calculations. What happens is that the solution for the fluctuations has two modes, the growing one and the decaying one. When the field bounces, fluctuations either grow or decay depending on the phase with which they re-enter the tachyonic regime. Therefore even though the modes with  $k \ll m$ grow fast on the way down, they also decay fast on the way up, as  shown in Fig. \ref{k=0}.

\begin{figure}[Fig001]
\centering \leavevmode\epsfysize=7cm
\epsfbox{\picdir{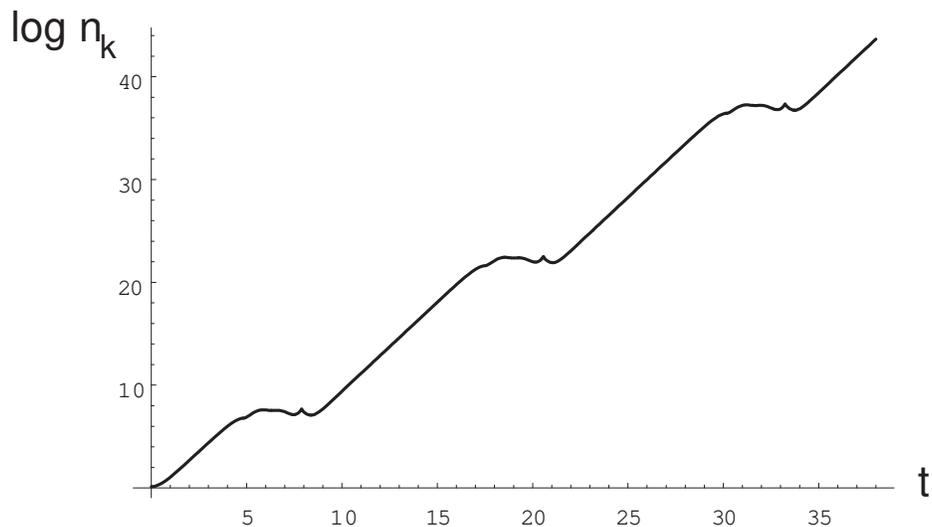}}

\

\caption[Fig001]{\label{k=05} {Same as in Fig 1
for  $k = 0.5 m$.}}
\end{figure}

Meanwhile for the modes in a rather broad interval of $k$, from $k \sim 0.3 m$ to $k \sim 0.6 m$, the modes continue their growth when the field oscillates.   Figure \ref{k=05}  shows the growth of $n_k$ during three consecutive oscillations of the field $\phi$. As we see, during each full oscillation the occupation numbers grow $e^{15}$ times. Thus during $n$ oscillations the occupation numbers should grow $e^{15 n}$ times.

This is an incredibly fast growth. It occurs much faster than the usual parametric resonance in theories with $m^2 > 0$ \cite{KLS}. Clearly, this process can rapidly convert all the energy of the homogeneous field into the energy of classical colliding waves, and at this stage the only reliable way to study the process is to use numerical simulations.

In the theory $-m^2\phi^2$ the fluctuations grow as fast as the scalar field, so if one begins with a homogeneous field with ${\delta\phi\over \phi_0} \ll 1$, then on the way down to $\phi \sim v$ this field distribution remains relatively homogeneous.
However, when the field rolls back towards $\phi = \phi_0$,  the inhomogeneities with $k \sim 0.5 m$ continue growing. For example, if one takes $\phi_0 \sim 10^{-2} v$, quantum fluctuations (which have initial amplitude only one order of magnitude smaller than $\phi_0$ in this model) grow almost $10^4$ times when the field $\phi$ falls down from $\phi = \phi_0$ and returns back. The amplitude of inhomogeneities after the return becomes approximately three orders of magnitude larger than $\phi_0$, which means that the homogeneity becomes completely destroyed. At this stage (and in fact much earlier) one can no longer study the evolution of quantum fluctuations as if they were small deviations on a homogeneous background. When the field falls down to the minimum of the effective potential again, it becomes divided into large colliding waves. One cannot study the evolution of such a system using  perturbation theory.

In the theories  $V(\phi) \sim -\phi^n$ with $n > 2$ the situation may be even more interesting and the growth of the occupation number $n_k$ for small $k$ occurs even faster. For example,  in the theory $-\lambda\phi^4$ long wavelength fluctuations grow as $\phi^2$ (and the occupation numbers grow as $\phi^4$). Therefore when the field $\phi$ grows from $\phi_0$ to $v$, the ratio ${\delta\phi\over \phi} $ grows by a factor of $O\left({v\over \phi_0}\right)$. This means that the field  may become very inhomogeneous on its way down even if initially it was very homogeneous.

The average initial amplitude of tachyonic fluctuations in the theory $-\lambda\phi^4/4$ at $\phi_0 \not = 0$ is given by $\delta \phi \sim {\sqrt{|V''|}\over 2\pi} \sim {\sqrt{3\lambda}\phi_0\over 2\pi}$. Initial level of inhomogeneities was given by ${\delta \phi \over  \phi} \sim  {\sqrt{3\lambda} \over 2\pi} \ll 1$. When the field $\phi$ reaches some value $v \gg \phi_0$, the ratio ${\delta \phi \over  \phi}$ grows  and becomes ${\delta \phi \over  \phi} \sim  {\sqrt{3\lambda} \over 2\pi} {v\over \phi_0}$. Thus, if the rolling of the field begins at a very small value of the field $\phi_0$, or if it continues long enough, so that ${v\over \phi_0}$ becomes greater than $1/\sqrt\lambda$, the field becomes completely inhomogeneous on its way down.

Moreover, if one considers a theory such as e.g. $V(\phi) = -\lambda \phi^4/4 + \lambda \phi^6/v^2$, which has a minimum at $\phi = v$, then in such theories, just as in the theory $-m^2\phi^2$, there are some modes with $k \sim \sqrt\lambda\phi_0$ whose amplitude grows both on the way down and on the way up. For these modes the degree of inhomogeneity rapidly grow with each oscillation. The
occupation numbers grow approximately as $\left({v\over\phi_0}\right)^8$ during each full oscillation, so that after $n$ oscillations the occupation numbers of the particles with momenta $\sim \sqrt\lambda\phi_0$ become as large as $\left({v\over\phi_0}\right)^{8n}$. That is why it takes only one or two oscillations before the  oscillating scalar field becomes inhomogeneous and the first stage of preheating related to the tachyonic instability completes.

The  simple rules derived above explain the extraordinary efficiency of tachyonic preheating. However, one can apply these rules only at the beginning of the process, when one can neglect the backreaction of created particles.    That is why we needed to perform computer simulations which took  the effects of backreaction into account.

\section{Lattice simulations of spontaneous symmetry breaking in the theories with  $V(\phi) = -m^2\phi^2/2 + \lambda \phi^4/4$}
\label{quadratic}

 A  description of our method \cite{lattice,FT} in application to this problem is given in the appendix.

\begin{figure}[Fig001]
\centering \leavevmode\epsfysize=14cm \epsfbox{\picdir{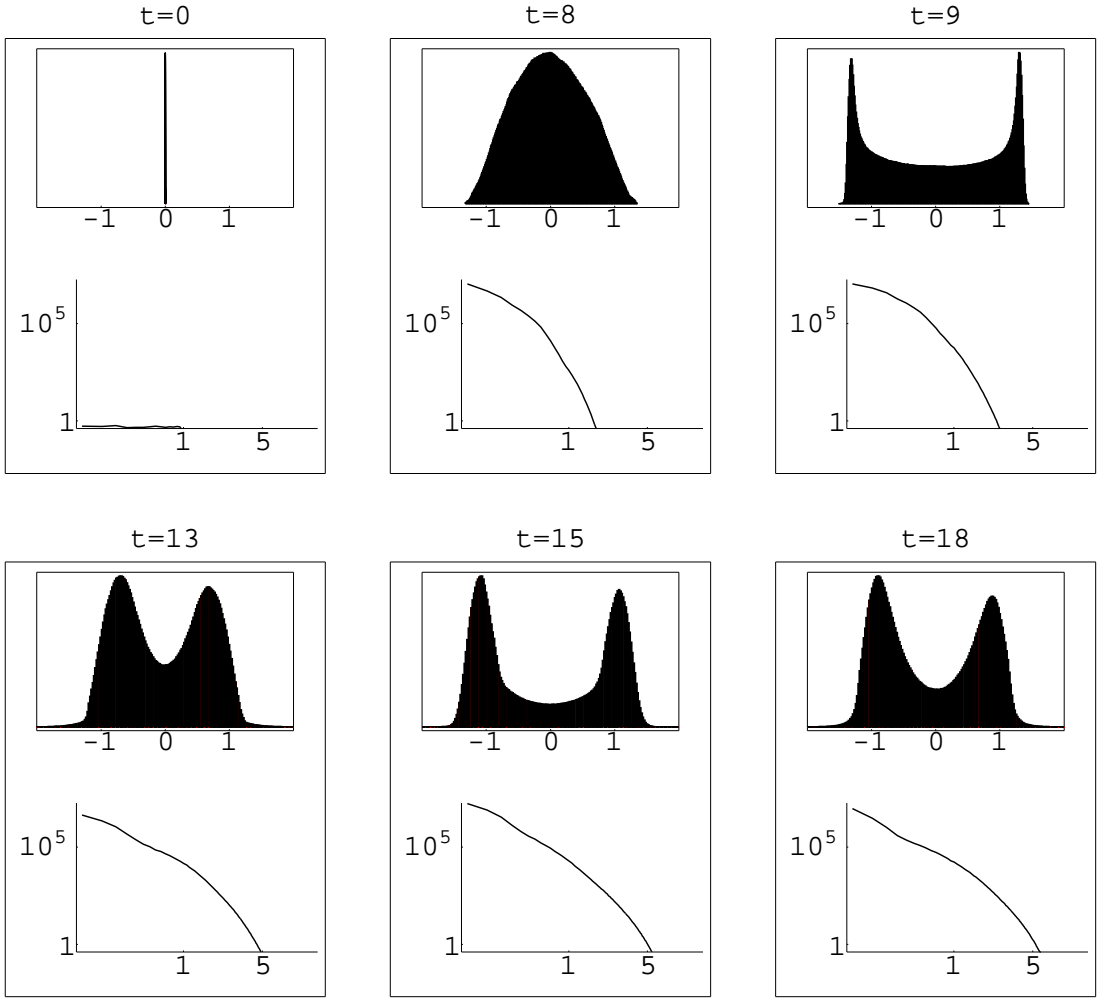}}

\

\caption[Fig001]{\label{onefielddistrib} The process of symmetry
breaking in the model (\ref{aB1}) for $\lambda = 10^{-4}$.  The values of the
field are shown in units of $v$, time is shown in units $m^{-1}$.  For each
moment of time, we also show the occupation numbers $n_k$ (the lower part of each panel), with $k$
measured in units of $m$.   At $t = 0$
one has  $n_k = 0$, as in the usual  quantum field theory vacuum. In the
 beginning of the process the occupation numbers $n_k$ grow exponentially for $k < m$ ($k < 1$ in the figure), but then this growth spreads to
$k > m$ because of domain wall formation and collisions of classical
waves of the field $\phi$. Within a single oscillation the occupation
numbers for $k \ll m$ grow up to  $\sim 10^6$, which is in complete agreement
with our estimate $n_k \sim 10^2 \lambda^{-1}$, Eq. (\ref{occlam}). The
spectrum rapidly stabilizes, but it is not thermal yet, and the
occupation numbers remain extremely large. Thermalization takes much more
time than spontaneous symmetry breaking.}
\end{figure}

There are several complementary ways one can represent the results of our
calculations. One of the best ways to do it is to study  the probability
distribution function $P(\phi,t)$, which is the fraction of the volume
containing the field $\phi$ at a time $t$.  At $t = 0$ we begin
with the probability distribution concentrated near $\phi = 0$,
with the quantum mechanical dispersion (\ref{aBB}), and then we follow its
evolution; see Fig. \ref{onefielddistrib}.

In the beginning   quantum fluctuations are very small, and the
probability distribution $P(\phi,t)$ is very narrowly focused near $\phi = 0$. Then it spreads out and shows
two maxima that oscillate about $\phi = \pm v$ with an amplitude much
smaller than $v$.

As we see from Fig. \ref{onefielddistrib}, the two maxima never come close to the initial point
$\phi = 0$, which implies that symmetry becomes broken within a single
oscillation of the distribution of the field $\phi$.  To demonstrate that this is not
a strong coupling effect, we show the results for the model
(\ref{aB1}) with $\lambda = 10^{-4}$.  We obtained similar results for $\lambda=10^{-2}$. Note that only when the
distribution stabilizes and the domains become large can one use
the standard language of perturbation theory describing scalar
particles as excitations on a (locally) homogeneous background.
That is why the use of the nonperturbative approach based on
lattice simulations was so important for our investigation.

One may wonder why the distribution is slightly asymmetric, and why  after symmetry breaking there
are still many points at $|\phi| \ll v$. The answer is that after
SSB space becomes divided into domains with $\phi \sim \pm v$.
Domains are large, and their size gradually grows after SSB because large domains ``eat'' the small  ones. Eventually in any finite size
box there will remain just one domain, i.e. the distribution will become
completely asymmetric.  The points with $|\phi| \ll v$ correspond to domain
walls.

In this series of simulations we made a cut-off in the spectrum of initial
fluctuations at $k > m$. The reason is that only the modes with $k < m$
from the very beginning experience exponential growth and behave as
classical fields. We checked, however, that the results of the simulations
remain qualitatively the same if one makes a cut-off at $k \gg m$.

The  process of thermalization takes  much longer than spontaneous symmetry
breaking  \cite{FK}. Indeed,  the standard thermal distribution is given by
the well known equation $
n_k=(e^{\omega_k / T} -1)^{-1} $. At the moment when all the energy $V(0) =
{m^4\over 4 \lambda}$ is transferred to the thermal energy $\sim T^4$, the
temperature rises up to $T \sim m \lambda^{-1/4}$, and the occupation
numbers at $k \lesssim m$ become $
n_k \sim (e^{m / T} -1)^{-1} \sim T/m \sim \lambda^{-1/4}$. In particular,
for $\lambda \sim 10^{-4}$ one would have $n_{k<m} = {\cal O}(10)$, which
is 5 orders of magnitude smaller than the results of our calculations.

Thus, the occupation numbers should drop down dramatically before full
thermalization is achieved. This may happen only if the total number of
particles becomes many orders of magnitude smaller (particle cannibalism).
If one considers only those interactions that preserve the total number of
particles (scattering $2 \to 2$), one may achieve a kind of temporary
thermal equilibrium with a nonvanishing effective chemical potential of
particles $\phi$. This is what we see in our calculations when the
occupation numbers gradually approach some quasi-equilibrium asymptotic
limit. Since there is no real particle conservation in this theory, eventually the effective chemical potential will vanish, and the true thermal equilibrium with $n_k=(e^{\omega_k / T} -1)^{-1} $ will be reached. But this process takes much greater time than the time required for spontaneous symmetry breaking.

\begin{figure}[Fig001]
 \centering
\leavevmode\epsfysize=21.5cm \epsfbox{\picdir{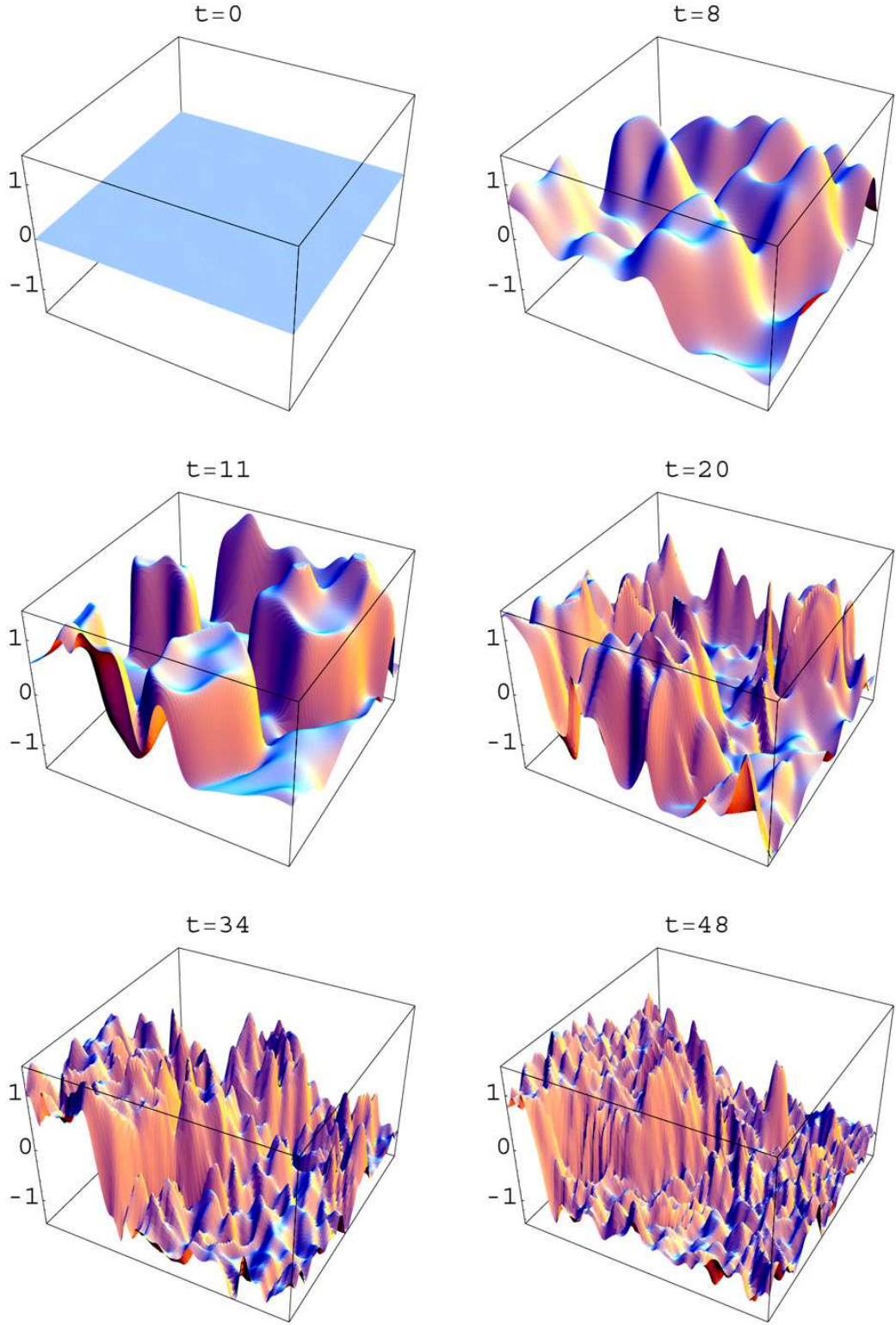}  }

\

\caption[Fig001]{\label{onefieldslice} {Tachyonic growth of quantum
fluctuations and the early stages of domain formation  in the simplest
theory of spontaneous symmetry breaking with $V(\phi)= - {m^2\over
2}\phi^2 + {\lambda\over 4} \phi^4$.  }}
\end{figure}

To provide a visual picture of the distribution of the scalar field, we show the growth of fluctuations in a
two-dimensional slice of 3D space in this model in Fig.
\ref{onefieldslice}.   Maxima correspond to domains
with $\phi >0$; minima correspond to domains with $\phi<0$. The third image
 corresponds to the first one half of an oscillation, just like the third
panel in Fig. \ref{onefielddistrib}. As we see the
universe at that moment is already divided into domains with $\phi \sim \pm v$. The initial  size
of each domain is somewhat greater than $m^{-1}$. Inside each domain the
deviation from  $\phi = \vert v\vert$ is much smaller than $v$.
This confirms our conclusion that spontaneous symmetry breaking occurs
within a single oscillation.

\newpage

\newpage

The original domain structure can change within a time
$O(10 m^{-1})$ because of domain wall collisions and domain expansion.
Gradually, the size of each domain grows and the domain wall structure becomes more and more stable, as we see in the last two images of Fig.
\ref{onefieldslice}.

\begin{figure}[Fig001]
\centering \leavevmode\epsfysize=15cm
\epsfbox{\picdir{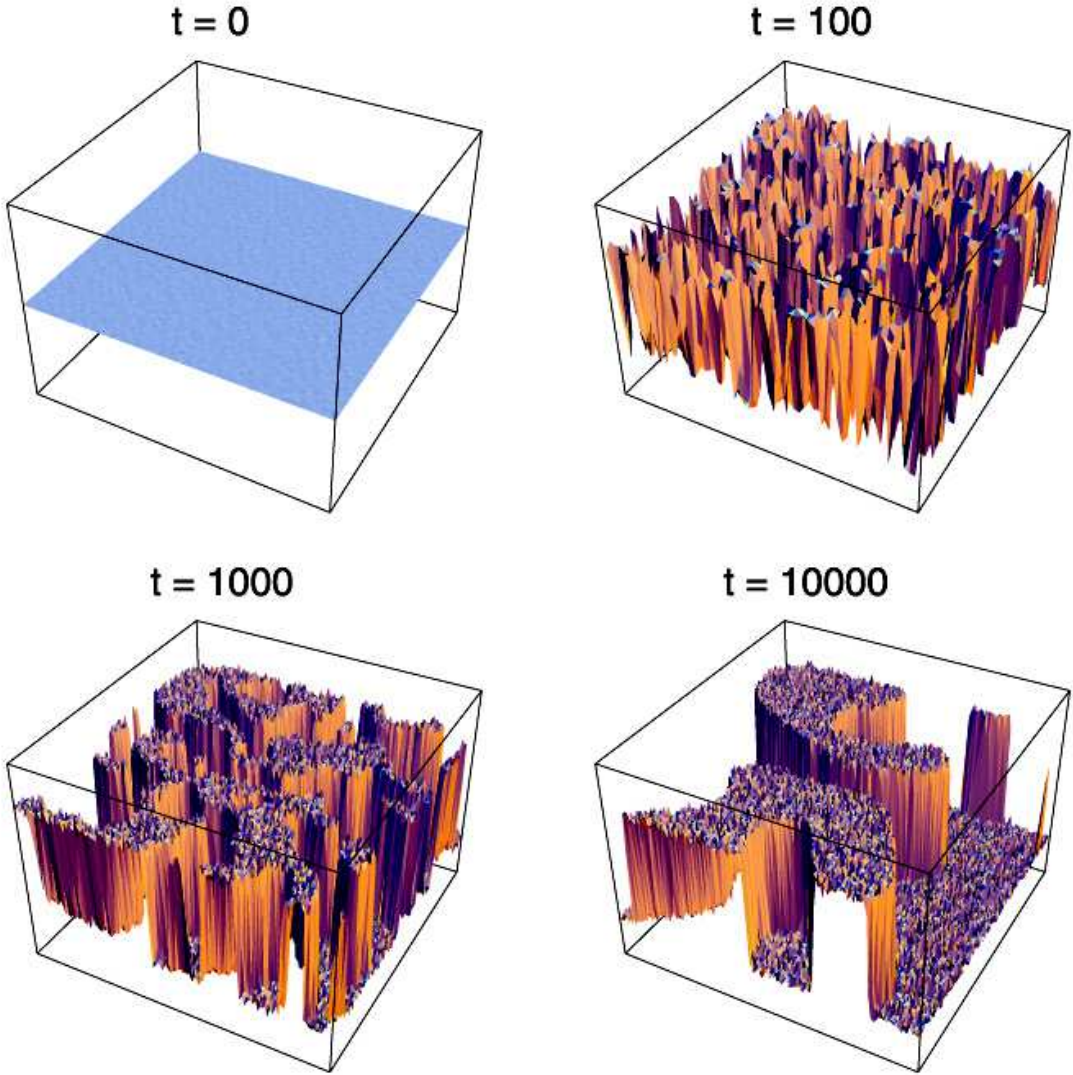}}

\

\caption[Fig001]{\label{domains} {Formation of domains in the process of
symmetry breaking in the  model (\ref{aB1}).  }}
\end{figure}

If one continues the calculation for a much longer time, one can see much more
clearly the formation and growth of domains with $\phi = \pm v$. In the
beginning these domains are small, but then they ``eat'' each other and
grow. To illustrate this process we performed simulations in a 2D box of
size 1024 x 1024. This allowed us to perform the calculations out to a much
greater time and see domain formation on a much greater scale;
see Fig. \ref{domains}. One should note that the process of rescattering
of particles produced during preheating in 2D is somewhat different from
that in 3D. However, the tachyonic instability is the same in both cases
and the process of domain growth is qualitatively similar.

\section{Comparison with the usual perturbative approach}\label{perturbative}

Before going any further, let us discuss the difference between our methods and the usual approach based on perturbation theory. In the usual approach one solves a self-consistent system of equations  for the homogeneous scalar field $\phi$ and for the variance $\langle\delta\phi^2\rangle$, where $\langle\delta\phi^2\rangle$ is generated due to particle production by the oscillating field $\phi$, see e.g. \cite{KLS,Boyan}.

In this approach one would expect that the effective mass of the field $\phi$ is given by $m^2_\phi = -m^2 + 3\lambda \langle\delta\phi^2\rangle$. This is the standard approach used, in particular, in the theory of the high temperature cosmological phase transitions \cite{book}. 

Let us see what would happen if we naively applied this method to our problem. Our investigation shows that within a single oscillation the variance of the scalar field  grows to $\langle\delta\phi^2\rangle \approx v^2$; see Fig. \ref{onefielddistrib}. A more detailed investigation shows that soon after the beginning of the process the value of $\langle\delta\phi^2\rangle$  is not much different from ${3\over 4} v^2$. This would seem to imply that immediately after spontaneous symmetry breaking  the symmetry becomes restored again, because the effective mass squared of the field $\phi$ becomes positive: $m^2_\phi \approx -m^2 +{9\over 4}\lambda v^2 \approx {5\over 4} m^2$.

However, our numerical calculations clearly demonstrate that this is not the case. So what is wrong with the standard perturbative approach?

The point is that in order to investigate the tachyonic instability one should study the local field distribution on a scale comparable to $m^{-1}$ instead of the field distribution averaged over the whole universe. In our scenario the universe becomes divided into different domains of size   greater than $O(m^{-1})$. If one wants to study the process of spontaneous symmetry breaking, then instead of finding the average values $\langle\phi \rangle$ and $\langle\delta\phi^2\rangle$ over the whole universe one should find the average value of $\phi$ inside each domain. After that,  one should calculate the  variance $\langle\delta\phi^2\rangle$, where $\delta\phi$ is the {\it local} deviation of the field $\phi$ from its average value inside each domain. But even this will give only partial information about the process. That is why in addition to finding the probability distribution and the occupation numbers  (Fig. \ref{onefielddistrib}),  we have shown the spatial distribution of the field $\phi$ (Fig. \ref{onefieldslice}).

 This need for local averaging is an important issue that was overlooked in many recent works on preheating, as well as in some works on the backreaction of long wavelength inflationary quantum fluctuations on the speed of expansion and the average energy-momentum tensor of matter. We are not saying that the calculation of averages such as $\langle\phi \rangle$ and $\langle\delta\phi^2\rangle$ over the whole universe is not useful. For example, it is quite informative in the theory of high temperature phase transitions, where the typical contribution to $\langle\delta\phi^2\rangle$ occurs due to the short wavelength fluctuations with the wavelength $T^{-1}$, which is much smaller than $m^{-1}$ at the time of the phase transition \cite{book}. However, one should be extremely careful using averages like $\langle\delta\phi^2\rangle$ over the whole universe in situations where a substantial contribution to $\langle\delta\phi^2\rangle$ is given by fluctuations whose wavelength is greater than the typical length scale of the problem. It does not matter how accurately one calculates such averages, whether one works in the Hartree approximation or in the $1/N$ approximation, as in \cite{Boyan}. In the case described above we calculated $\langle\delta\phi^2\rangle$ very accurately using our lattice simulations. This method takes into account not only the usual backreaction effects that could be studied in the Hartree or $1/N$ approximations, but also effects of rescattering of produced particles. Still we have seen that a naive use of our results would lead to an incorrect conclusion that symmetry becomes restored immediately after it breaks down.

If one has to use perturbation theory in situations when the infrared contribution to $\langle\delta\phi^2\rangle$ is substantial, the occupation numbers are large and the results allow a semi-classical interpretation, one can avoid the problem discussed above if one rearranges perturbation theory in a nontrivial way. For example, if one studies effects on a length scale $l$, one may consider all fluctuations on  larger scales as a   nearly homogeneous classical field background and ignore the  contribution of these fluctuations to $\langle\delta\phi^2\rangle$. In the context of the theory of preheating, this issue was discussed in Section X of Ref. \cite{KLS}.  In inflationary cosmology a similar approximation constitutes the basis of the stochastic approach to inflation \cite{Star,LLM,book}.  An alternative method is to use numerical simulations that can bring us more detailed   information about the process. This is the method we use in our paper.

To compare our method with the more traditional perturbative approach assuming initial displacement of the field $\phi_0 \not = 0$, we performed a series of simulations for different values of $\phi_0$ exceeding the level of the long wavelength quantum fluctuations $\sim {m\over 2\pi}$. For $\phi_0 \approx 10^{-2} v$ the results did not differ much from the results that we obtained for $\phi_0 = 0$ in the previous section. The distribution of the field $\phi$ never returned back to the vicinity of $\phi = 0$, and the process of spontaneous symmetry breaking occurred within a single oscillation. Note, that our calculations were performed for $\lambda \sim 10^{-4}$, so that $\phi_0 \approx 10^{-2} v \gg {m\over 2\pi} \sim {10^{-2} v\over 2\pi}$.

\begin{figure}[Fig001]
\centering \leavevmode\epsfysize=20cm \epsfbox{\picdir{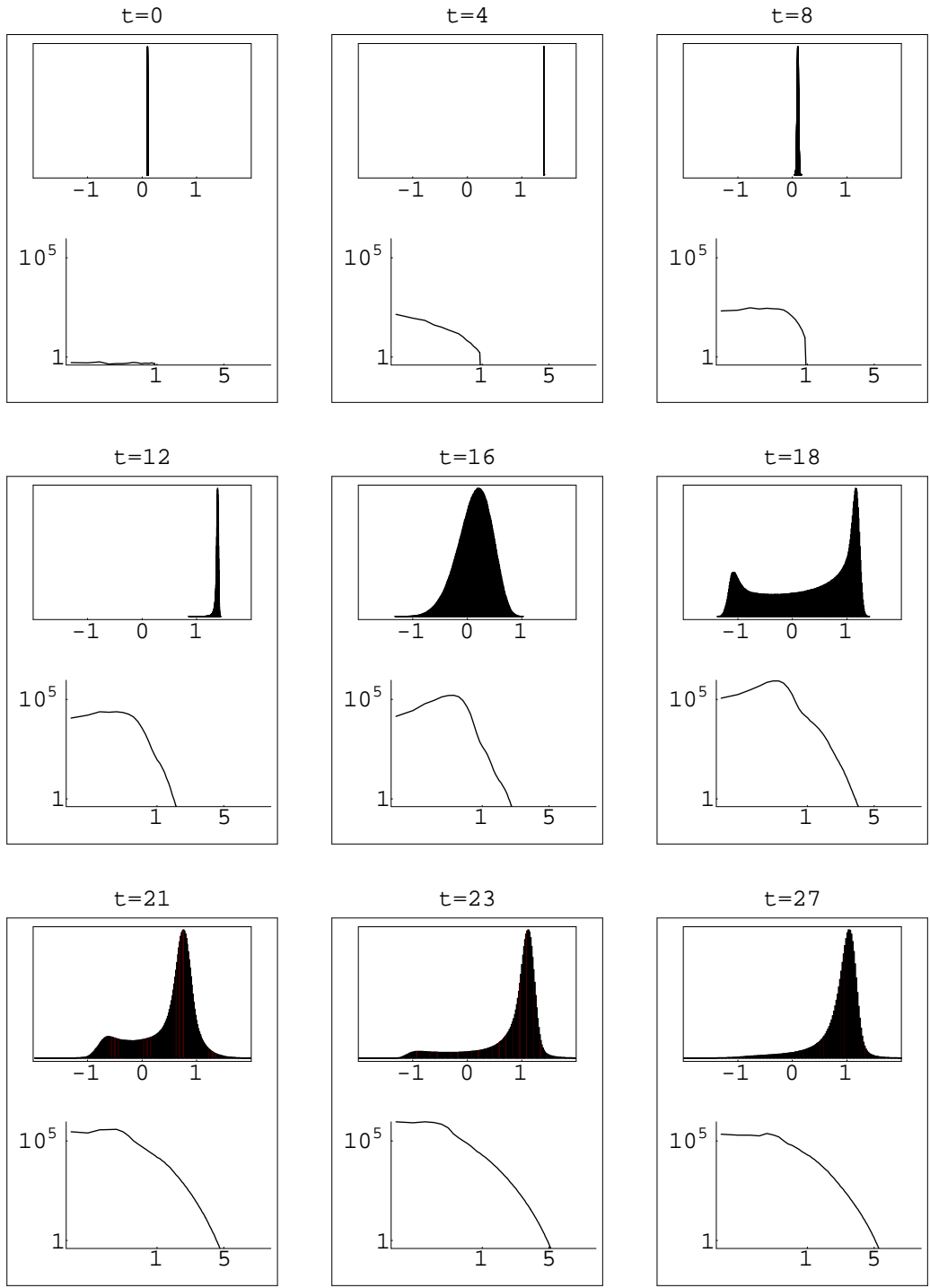}}

\

\caption[Fig001]{\label{onefielddistribdispl} The process of symmetry
breaking in the model (\ref{aB1}) for $\lambda = 10^{-4}$ in the case where
the field $\phi$ was initially displaced from $\phi = 0$ by $\phi_0 = 10^{-1} v$. As we see, in this case spontaneous symmetry breaking takes two oscillations to occur, and the occupation numbers are smaller than in the case of falling from $\phi = 0$. }
\end{figure}

For a much greater initial displacement the results were somewhat different, but still for $\phi_0 \ll v$ we have found that the regime of homogeneous oscillations completely disappeared after a couple of oscillations.
Consider, for example, the case $\phi_0 = 0.1 v$; see Fig.
\ref{onefielddistribdispl}. As one might expect, in this case the final probability distribution
is entirely concentrated at $\phi \sim +v$, and one can check that no topological defects are
produced at the end of the process. But even in this case we have found that the process completes very fast, after the second oscillation.

This result  differs from the results of  investigation of the same model in \cite{Boyan}, where it was claimed that the oscillations of the homogeneous component of the field in this model continue for a long time with the amplitude comparable to $v$, even if one starts with $\phi_0 \ll {m \over 2\pi}$. The reason for the disagreement is very simple. First of all, in the  investigation of spontaneous symmetry breaking in the theory $   - {m^2\over 2}\phi^2 + {\lambda\over 4} \phi^4 + {m^4\over
4\lambda}$ in \cite{Boyan}  the initial displacement of the field $\phi_0$ was chosen two orders of magnitude smaller than the level of the long wavelength quantum fluctuations with $k < m$, $\delta\phi \sim {m\over 2\pi}$. In this case SSB appears not because of the growth of the homogeneous component of the field, but because of the generation of fluctuations with $k < m$. In such a situation investigation of the homogeneous component of the field does not give much information about spontaneous symmetry breaking.

Moreover, as soon as the combined amplitude of all fluctuations with $k < m$ (i.e. $\sqrt{\langle\delta\phi^2\rangle_{k < m}}$) becomes comparable to  $v$, which happens much earlier than the homogeneous component of the field reaches the minimum of the effective potential, the universe becomes divided into domains with colliding walls. At this moment the standard perturbative approach completely breaks down. It does not describe rescattering of produced particles, collisions of classical waves of the scalar field, and dynamics of topological defects. In this regime equations describing the evolution of the homogeneous component of the field $\phi$ derived in the Hartree approximation (or in 1/N approximation) become inapplicable. That is why in our work we studied SSB using a combination of analytical investigation and lattice simulations.


\newpage

\section{Spontaneous symmetry breaking in the theory of a complex
field}\label{complexsection}

\begin{figure}[Fig001]
\centering \leavevmode\epsfysize=14cm
\epsfbox{\picdir{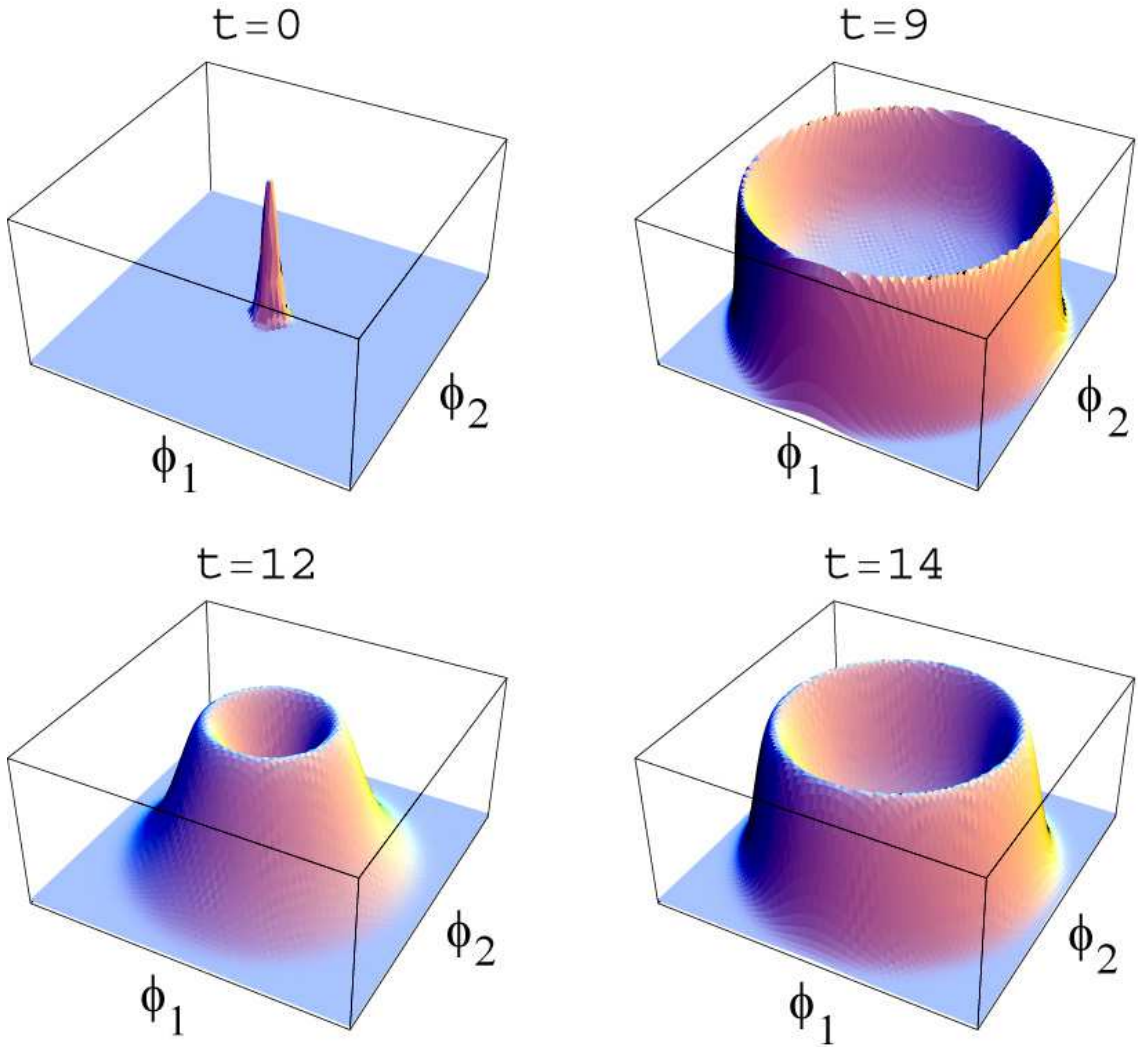}}

\

\caption[Fig001]{\label{complex} {The process of symmetry breaking in the
model (\ref{aB1}) for a complex field $\phi = {1\over \sqrt 2} (\phi_1 + i
\phi_2)$. The field distribution falls down to the minimum of the effective
potential at $|\phi| = v$ and experiences only small oscillations with
rapidly decreasing amplitude $|\Delta\phi| \ll v$.}}
\end{figure}

In the previous section we studied symmetry breaking in a theory (\ref{aB1}) describing  a one-component real field $\phi$.
One can perform a similar investigation for the theory of a multi-component
scalar field $\phi_i$ with the potential (\ref{aB1}), simply replacing $\phi^2$ with $\vert\phi\vert^2$.  Figure \ref{complex}
  illustrates the dynamics of symmetry breaking
in the model (\ref{aB1}) with a two-component scalar field $\phi =
(\phi_1 + i\phi_2)/\sqrt 2$. It shows the probability distribution
$P(\phi_i,t)$, which is the fraction of the volume containing the
field $\phi$ at a time $t$.

In the beginning, the
probability distribution is concentrated near $\phi = 0$, with the
quantum mechanical dispersion (\ref{aBB}). Then the probability distribution spreads out, and after a single
oscillation it  stabilizes at $|\phi|
\sim v$, which corresponds to SSB.
The standard approximation representing the scalar field
as a homogeneous background field with small fluctuations does not
work at any stage of the process.

A detailed investigation of the spatial distribution of the field $\phi$
shows \cite{GBFKLT} that after the first oscillation the scalar field can
be represented as a collection of classical waves oscillating near $|\phi|
\sim v$ with an amplitude smaller than $v/2$. Thus SSB indeed occurs within a single oscillation of the field distribution. A small but nonvanishing
height of the histogram in Fig. \ref{complex} at $\phi = 0$ is due to the
presence of strings that  have $\phi = 0$ at their cores.

\begin{figure}[Fig001]
\centering \leavevmode\epsfysize=8cm \epsfbox{\picdir{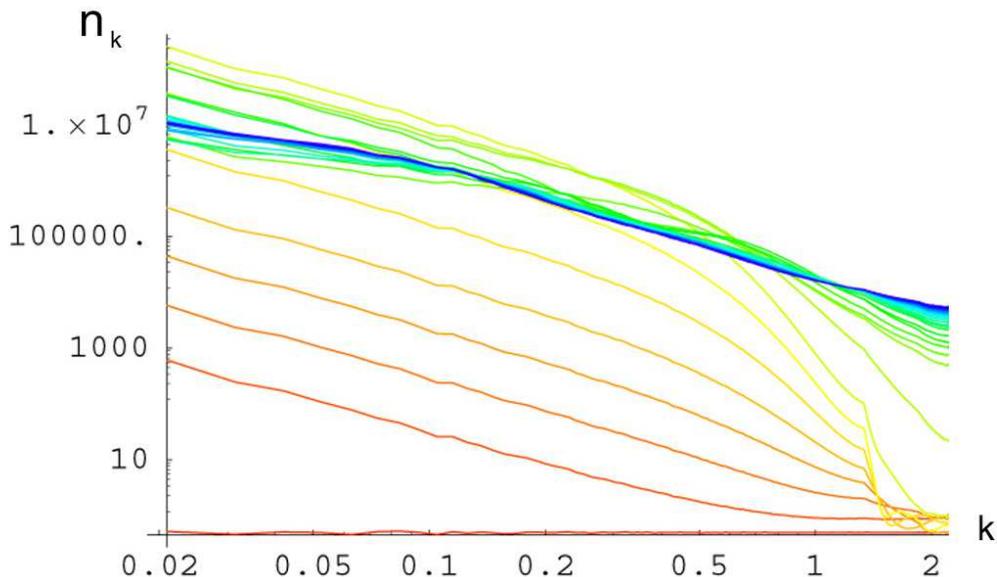}}

\

\caption[Fig001]{\label{compocc} {Occupation numbers $n_k$ of particles
produced during tachyonic preheating in the model of a complex scalar field
$\phi$ with effective potential $V=-  m^2  \phi^*\phi +  \lambda
 (\phi^*\phi)^2$ with $\lambda=10^{-4}$. In the beginning (lower curves), $n_k$ grows for
$k \lesssim m$ ($k \lesssim 1$ in this figure), but then eventually this
growth spreads to larger $k$.}}
\end{figure}

Fig. \ref{compocc} shows the occupation numbers $n_k$ of produced
particles. During the first oscillation these numbers grow up to $10^7$  --
  $10^8$ for $k < m$ ($k < 1$ in the figure).
Then the occupation numbers at  $k < m$ slightly decrease, whereas the
occupation numbers at $k > m$ begin to grow. Complete thermalization
takes a very long time.

\begin{figure}[Fig001]
\centering \leavevmode\epsfysize=11cm \epsfbox{\picdir{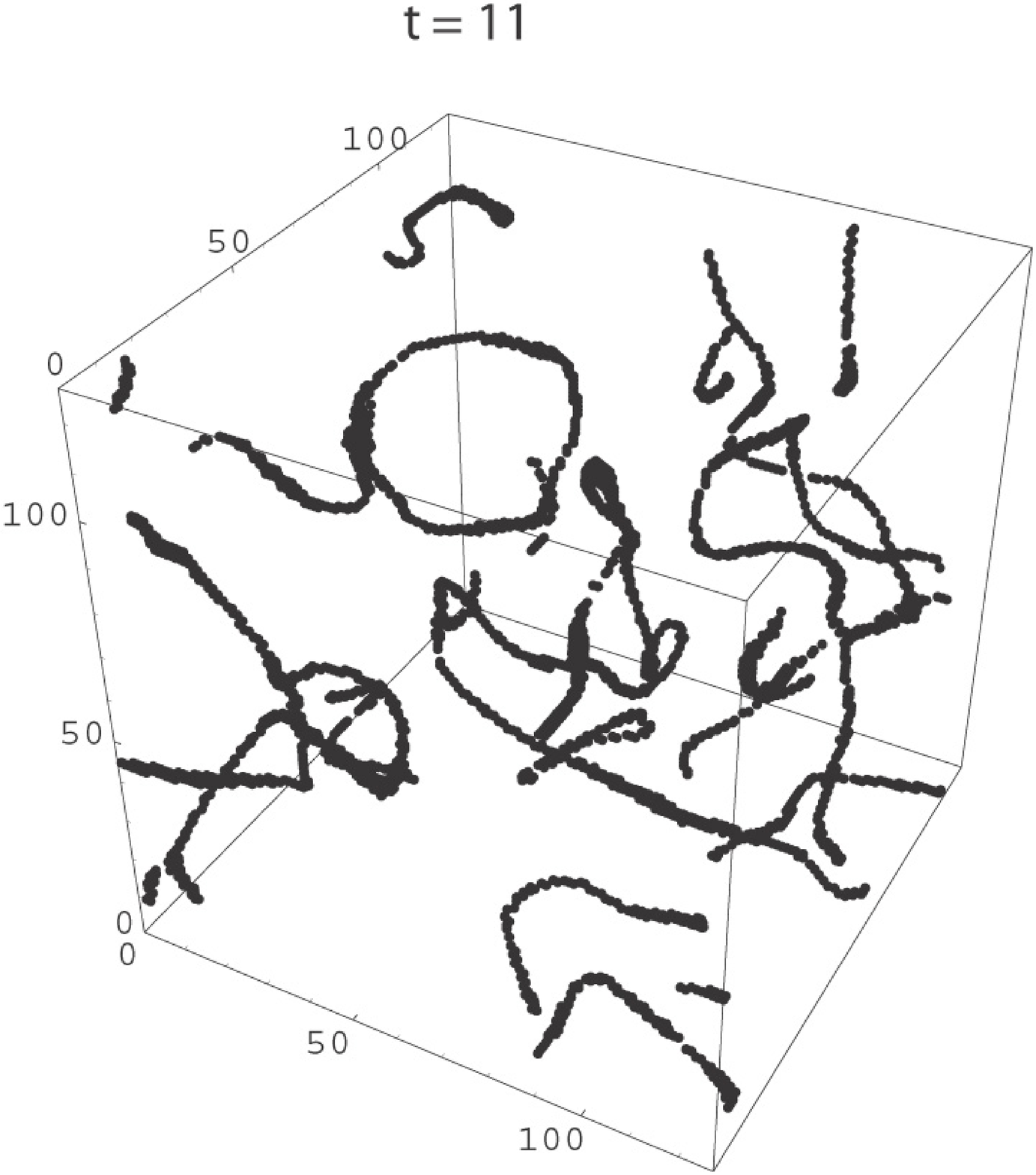}}

\

\caption[Fig001]{\label{strings} {Strings produced after  one half
of an oscillation in the model (\ref{aB1}) for a complex field $\phi$.}}
\end{figure}

In the model of a complex scalar field, instead of domain walls one has
strings that are produced when the field falls down; see Fig. \ref{strings}.  The whole process of
string formation occurs within a single oscillation.  After that the new
long strings are not formed. Sometimes small string loops appear and
disappear because of occasional large fluctuations of the scalar field.  If
symmetry were broken and then restored again when the field distribution
moves back to $\phi = 0$, we would see strings being ``melted,'' and then a completely new set of strings would appear. Meanwhile, our simulations show that the large scale string distribution is formed as the field $\phi$ first rolls down to the minimum of the effective potential. During the subsequent oscillations the strings formed in the beginning of the process  do not disappear and are not replaced by  new ones; instead they experience only gradual evolution. This confirms our conclusion that symmetry breaking is achieved within a single oscillation.

Just as in the case of the one-component scalar field, perturbative methods of investigation of this theory cannot describe formation of topological defects and scattering of classical waves produced by the tachyonic instability. Therefore such methods break down within the first oscillation of the field distribution.

\section{Quartic potential}\label{quarticsection}

The process of SSB will occur in a somewhat different way in theories
where the curvature of the effective potential near its maximum depends on
$\phi$. For example, one may consider the Coleman-Weinberg model, which was the
basis for the first version of the new inflation scenario \cite{New}:
\begin{equation}\label{CW}
V =  {\lambda\over 4} \left(\phi^4\log {\phi^2\over v^2}- {\phi^4\over2} +
{v^4\over2} \right).
\end{equation}
This potential has a maximum at $\phi = 0$ and a minimum  at $\phi = v$. At
small $\phi$ the effective potential looks like  $-{\lambda\phi^4\over 4}$
with an   effective coupling constant $\lambda(\phi) = \lambda \log
{v^2\over \phi^2}$.

Another interesting example is the toy model
\begin{equation}\label{CW2}
V =  -{\lambda\over 4} \phi^4+  {\alpha\phi^6\over 6} + {v^4\over 12}.
\end{equation}

An important feature of such potentials is that the tachyonic mass
$m^2(\phi) = V''(\phi)$ vanishes at $\phi = 0$. Therefore the simple
arguments based on the tachyonic growth of small quantum fluctuations do
not apply here. The decay of the symmetric phase in such models occurs via
tunneling and the
formation of bubbles. Historically, this was the first example of a theory
where tunneling occurs between two states ($\phi = 0$ and $\phi \not = 0 $)
even though there is no barrier separating these states \cite{Linde1}; see
also \cite{LW}.

To study symmetry breaking in these models one should first consider the
growth of the field $\phi$ in the model
\begin{equation}\label{quartic}
V =  -{\lambda\over 4} \phi^4   \ ,
\end{equation}
and then see what happens when one adds extra terms that stabilize the
potential.

The tunneling trajectories (instantons) with minimal action
 possess the $O(4)$
symmetry of  Euclidean space \cite{Coleman}. The Euclidean equation for
 O(4)
symmetric tunneling is
\begin{equation}\label{2}
 \phi''  +  3  \phi' r^{-1} = V'(\phi) \  .
\end{equation}
with the boundary conditions $\phi(r=\infty) = v$ and
$\phi'(0) = 0$. Here $\phi'(r) = {d\phi\over d r}$, $r = \sqrt {x^2_i}$;
the $x_i$ are the
Euclidean coordinates, i = 1,2,3,4.

Equation (\ref{2}) in the theory (\ref{quartic})    has a family
 of solutions \cite{Fub,Linde1}
\begin{equation}\label{3}
\phi(r) = 2\sqrt {2\over \lambda}
\left({\rho\over{r^2 + \rho^2}}\right) \ ,
\end{equation}
where $\rho$ is arbitrary.
Note that the value of the scalar field in the center of the bubble depends on $\rho$:
\begin{equation}\label{3a}
\phi(0) = {2\sqrt 2 \over \sqrt \lambda \rho}  \ .
\end{equation}
The corresponding Euclidean action
 does not depend
on $\rho$,
\begin{equation}\label{4}
S_E = 2\pi^2 \int r^3 \left({1\over 2} (\phi')^2 + V(\phi)\right) dr =
{8 \pi^2\over 3\lambda} \ .
\end{equation}
The
probability of bubble formation per unit four-volume can
 be estimated by
the expression
\begin{equation}\label{5}
P \sim \rho^{-4} \exp {\left(-{8\pi^2\over 3\lambda}\right)}  \sim \lambda^2\phi^4(0) \exp {\left(-{8\pi^2\over 3\lambda}\right)} .
\end{equation}

The probability of tunneling in the Coleman-Weinberg theory (\ref{CW}) can
be estimated by this equation if instead of $\lambda$ one uses the
effective coupling constant $\lambda \log{v^2\over \phi^2}$. Tunneling is
not strongly suppressed at
$\lambda \log{v^2\over \phi^2} \sim 1$. This means that tunneling
occurs to a point with exponentially small $\phi$: \ $\phi \sim v\
e^{-C/\lambda}$, with $C = O(1)$.

On the other hand, in the model (\ref{CW2}) the effective coupling constant $\lambda$ and  the  factor $\exp {\left(-{8\pi^2\over 3\lambda}\right)}$ suppressing  the tunneling do  not depend on $\phi$, whereas the subexponential factor in the expression for the tunneling probability (\ref{5}) is greater for large $\phi$. Thus in this model tunneling may occur to relatively large $\phi$.

The bubbles that appear after the tunneling are  described by eq.
(\ref{3}) if one understands by $r^2$ its Minkowski counterpart ${\bf r}^2
- t^2$:
\begin{equation}\label{3aa}
\phi(r) = 2\sqrt {2\over \lambda} \left({\rho\over{r^2-t^2 +
\rho^2}}\right) \ .
\end{equation}
Such bubbles have symmetry $O(3,1)$. When the bubble appears (at $t = 0$), the field takes its maximal value
$\phi_0$ at the center of the bubble, $\phi_0 = {2\over \rho}\sqrt {2\over
\lambda}$. Then it grows, and becomes infinitely large at $t = \rho =
{2\over \phi_0}\sqrt {2\over \lambda}= {2\sqrt{6}\over m_\phi(\phi_0)}$.
Here $m^2_\phi(\phi) =  V''(\phi) = 3\lambda\phi^2$.

Of course, in realistic models like (\ref{CW}) and (\ref{CW2}) the
field does not grow indefinitely large. It reaches the minimum of the
effective potential at $\phi = \pm v$ and then it begins oscillating
there.  Meanwhile quantum fluctuations may grow on top of the smooth instanton solution. The investigation of these oscillations and bubble wall collisions is a
complicated problem that can be studied numerically. Fortunately, the
behavior of the oscillating field prior to the bubble wall collisions  and neglecting quantum fluctuations can
be studied analytically by making a certain change of variables.

Indeed, it is known  that in properly chosen coordinates the interior of
each bubble looks like an open universe filled by a  {\it homogeneous}
scalar field $\phi$ \cite{ColDL}. One can show that during the main part of
the first oscillation of the field the radius of curvature (scale factor)
of this open universe is $O(\rho)$, which leads to expansion of the open
universe with Hubble constant $H \sim \rho^{-1} \sim
m_\phi(\phi_0)$.\footnote{In this paper we are neglecting the
overall expansion of the universe caused by the energy density of the
scalar field. This effect will be considered in a separate publication.
Here we consider ``expansion'' as it is seen inside the bubble in the
coordinate system in which the interior of the bubble looks like a
homogeneous open universe. This is not a physical expansion but a consequence of the choice of the coordinate system  in flat space where it is more convenient to study the bubble motion.}
 This introduces the damping term $3H\dot\phi$ to the equation of motion of
the scalar field, which gradually diminishes the amplitude of its
oscillations. Suppose that the tunneling occurs to $\phi_0 \ll v$, as in
the Coleman-Weinberg model. Then  during the main part of the first
oscillation  the effective mass of the field $\phi$ remains much greater
than $H \sim m_\phi(\phi_0)$, so in the limit ${\phi_0\over v} \to 0$ one
can neglect the effect of expansion of the open universe on the amplitude  of the oscillations. Later on, the Hubble constant in the open universe
bubble becomes even smaller and its damping effect on the oscillations
becomes even less significant. As a result, the amplitude of oscillations
of a homogeneous scalar field for a long time remains almost unchanged.

From the point of view of an outside observer using the usual coordinates
$x$ and $t$ this means that the field $\phi$ in the center of the bubble
oscillates for a long time with amplitude $O(v)$, sending spherical
waves of the same amplitude in all directions. The amplitude of each wave
is a function of $x^2-t^2$, which means that they propagate with a speed
asymptotically approaching the speed of light and their amplitude does not
depend on their distance from the center of the bubble.\footnote{One could
speculate about the possibility of sending a signal to aliens using such waves
with an amplitude that does not decrease with the distance. The  problem is
that these waves are only possible as a result of vacuum decay, which  first kills those who send the signal and then those who receive it.}

\begin{figure}[Fig001]
\centering \leavevmode\epsfysize=10.5cm \epsfbox{\picdir{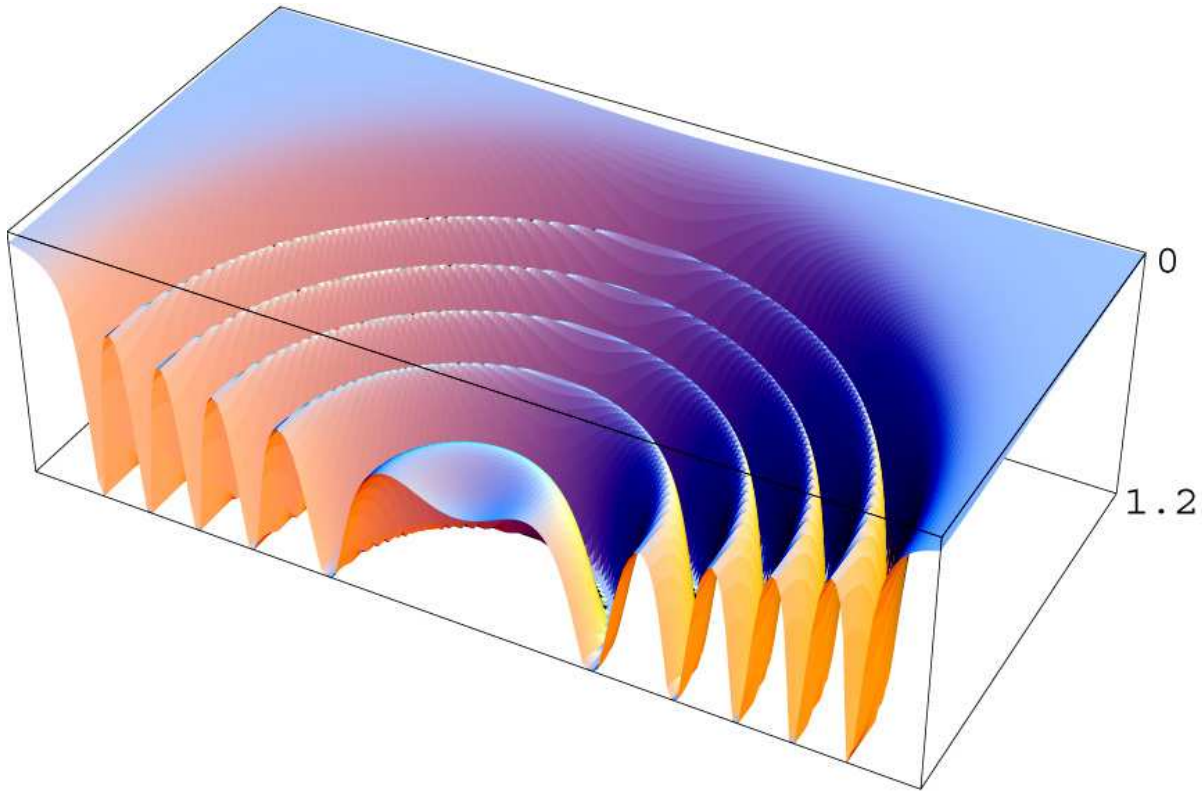}}

\

\caption[Fig001]{\label{instJPEG}  Field values on a partial 2D slice through the lattice in the model $V =  {\lambda\over 4} \left(\phi^4\log {\phi^2\over
v^2}- {\phi^4\over2} + {v^4\over2} \right)$. The process of symmetry breaking occurs due to tunneling and
bubble formation. After the tunneling, the bubble grows, and the field
inside it begins to oscillate. If the tunneling occurs from $\phi = 0$ to
$\phi_0 \ll v$, the amplitude of oscillations remains large for a long
time, and instead of the usual picture of a single bubble wall propagating
in all directions one has a series of propagating  waves with amplitude
comparable to $v$. The figure shows a half of such bubble, which appears
after the tunneling to  $\phi_0 = .02 v$. Cutting the bubble in half
allows us to see that the amplitude of oscillations  decreases rather
slowly, just as we expected.  Because the bubbles in this model take an exponentially long time to form, in our simulations we did not start at $\phi=0$ and wait for one to appear, but rather
started using the analytic form of the instanton as our initial
conditions. }
\end{figure}

Thus instead of the naive picture of a bubble  consisting of a single
spherically symmetric shell (which would be a correct picture in the
thin-wall approximation), one has a series of waves of almost equal
amplitude following each other; see Fig. \ref{instJPEG}. Reheating in this
model occurs due to a combination of different effects. First of all,
particles are produced during the collision of waves produced by different
tunneling events. But they are also produced due to the tachyonic
instability, as well as by the oscillations of the scalar field inside each
bubble.

\begin{figure}[Fig001]
\centering \leavevmode\epsfysize=12cm \epsfbox{\picdir{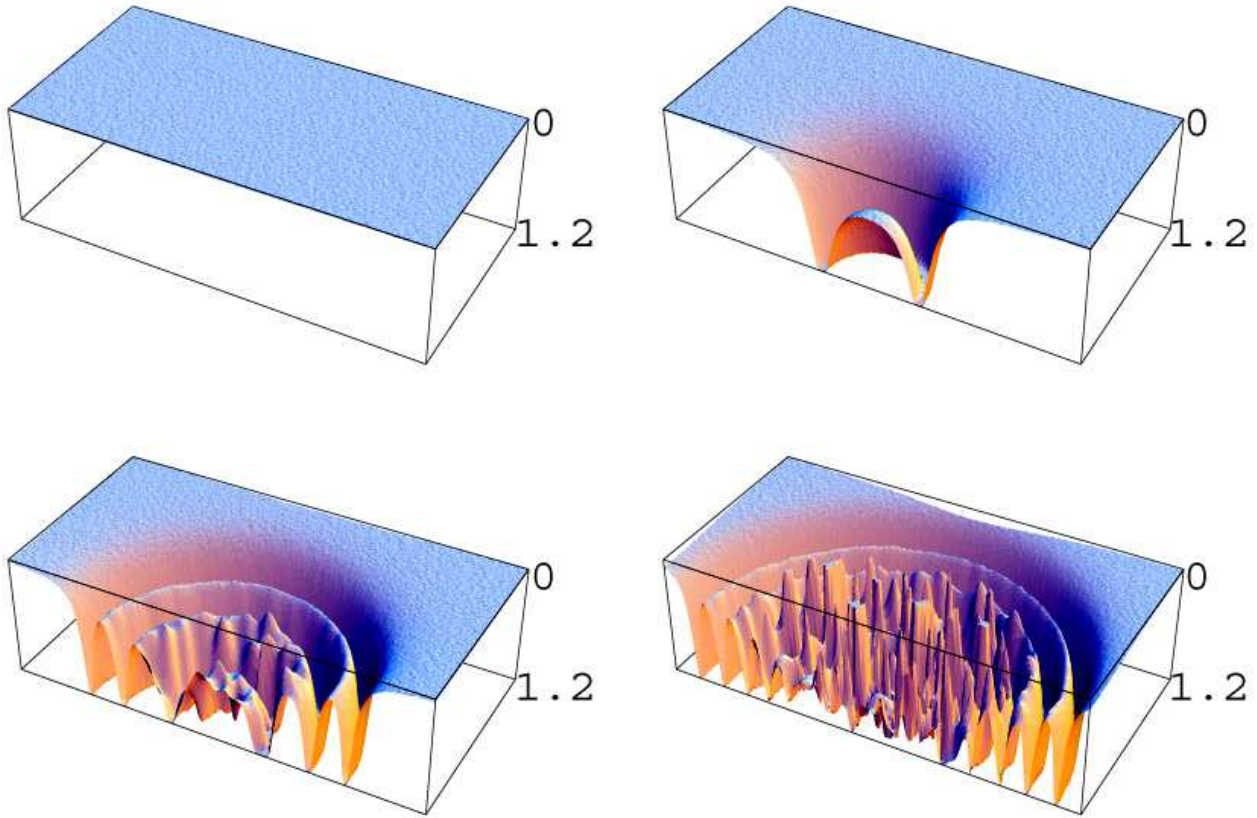}}

\

\caption[Fig001]{\label{instJPEG2} The process of symmetry
breaking in the model $V =  {\lambda\over 4} \left(\phi^4\log {\phi^2\over
v^2}- {\phi^4\over2} + {v^4\over2} \right)$ taking into account quantum fluctuations in the instanton background for $\lambda \sim 10^{-4}$.  As we see, quantum fluctuations lead to a growing asymmetry and decoherence of the oscillations due to the tachyonic preheating inside the bubble. Thus preheating in this model occurs due to a combined effect of bubble wall collision and tachyonic preheating. The latter mechanism is especially efficient if the tunneling occurs to $\phi_0 \ll v$. }
\end{figure}

We should make some comments here. First of all, if the
tunneling occurs to very small values of $\phi$, quantum fluctuations produced due to tachyonic instability inside the $O(3,1)$ symmetric bubble may
completely  distort the shape of the bubble during the field oscillations.  Within few oscillations, tachyonic preheating creates colliding waves inside the bubble; see Fig. \ref{instJPEG2}.

Note that each tunneling event produces an exponentially large  sphere
filled either by a positive field $\phi$ oscillating  around $\phi = v$ with a
slowly decreasing amplitude, or by a negative field $\phi$, oscillating
around $\phi = - v$. In both cases SSB occurs within a single oscillation
within each bubble, and then finally the field $\phi$ relaxes near $\pm v$ due to a combined effect of the amplitude decrease because of the bubble expansion, and the development of tachyonic preheating, as in Fig. \ref{instJPEG2}.

Finally, we should say that if the tunneling occurs to extremely small values of $\phi$, or if it does not occur for a long time, one may obtain an inflationary regime \cite{New}. Tachyonic preheating in this regime will be discussed in a subsequent
publication \cite{FKL2}.

\section{Cubic potential and stochastic approach to tunneling}\label{cubic}

Another important example of tachyonic preheating is provided by the
theory
\begin{equation}
V= -{\lambda\over 3} v\phi^3+{\lambda\over 4}\phi^4+{\lambda\over 12} v^4 \
.
\label{cub}
\end{equation}
This potential is a prototype of the potential that appears in
descriptions of symmetry breaking in F-term hybrid inflation
\cite{Stewart,hybrid2}.

The development of instability in this theory presents us with a new challenge.
The curvature of the effective potential at $\phi = 0$ in this theory
vanishes, which means that, unlike in the theory $-m^2\phi^2$ (\ref{aB1}),
infinitesimally small perturbations in this theory do not grow. On the
other hand, unlike in the theory $-\lambda\phi^4$
(\ref{quartic}), there are no instantons in this theory that would describe tunneling
from $\phi = 0$. Thus, in the theory $-{\lambda } v\phi^3$, which occupies
an  intermediate position between $-m^2\phi^2$ and $-\lambda\phi^4$, both
mechanisms that could lead to the development of instability do not work.
Does this mean that the state $\phi = 0$ in this theory is, in fact, stable?

The answer to this question is no; the state $\phi = 0$ in the theory
$-{\lambda } v\phi^3$ is unstable. Indeed, even though $\langle\phi\rangle$
initially is zero, long wavelength fluctuations  of the field $\phi$ are
present, and they may play the same role as the homogeneous field $\phi$ in
triggering the instability.

 Eq. (\ref{aBB}) implies that scalar field
fluctuations with momenta $k \lesssim k_0$  have initial amplitude
$\langle \delta\phi^2 \rangle \sim {k_0^2\over 8\pi^2}$. Thus the short
wavelength fluctuations with momenta $k > k_0$ live on top of a long
wavelength field with an average amplitude $\delta\phi_{\rm
rms}(k_0) \sim \sqrt {\langle \delta\phi^2 \rangle} \sim {k_0 \over 2 \sqrt
2 \pi }$.

The curvature of the effective potential $V'' = |m^2_{\rm eff}|$ at $\phi
\sim \delta\phi_{\rm
rms}(k_0)$ in the theory (\ref{cub}) is given by $-2\lambda v
\delta\phi_{\rm
rms}(k_0) \sim - \lambda v  {k_0 \over \sqrt 2 \pi }$. Consider
fluctuations with momentum $k$  somewhat greater than $k_0$, so that the
amplitude of the long wavelength field $\delta\phi$ does not change
significantly on a scale $k^{-1}$. Short wavelength fluctuations with $k =
C k_0$ with $C$ somewhat greater than $1$ will grow on top of the field
$\phi \sim \delta\phi_{\rm
rms}(k_0)$ if $k^2 \lesssim |m^2_{\rm eff}|   \sim {\lambda v  k_0
\over \sqrt 2\pi} $.

Taking for definiteness $C \gtrsim \sqrt 2$, one may argue that
fluctuations with $k \lesssim {\lambda v \over 2\pi}$ may enter a
self-sustained regime of tachyonic growth. Small fluctuations rapidly grow
large, which  justifies using semi-classical methods for the description of
this process. The average initial amplitude of the growing tachyonic
fluctuations with
momenta smaller than ${\lambda v \over 2\pi}$ is
\begin{equation}\label{typical}
 \delta\phi_{\rm rms} \sim {\lambda v   \over 4\pi^2}.
\end{equation}
These fluctuations grow until the amplitude of $\delta\phi$
becomes comparable to $2v/3$, and the effective tachyonic mass
vanishes. At that moment the field can be represented as a
collection of waves with dispersion $\sqrt{\langle
\delta\phi^2\rangle} \sim v$, corresponding to coherent states of
scalar particles with occupation numbers $n_k \sim
\left({4\pi^2\over \lambda}\right)^2 \gg 1.$   

A more accurate investigation
shows that the initial value of the field is few times greater than
$\delta\phi_{\rm rms} \sim {\lambda v   \over 4\pi^2}$ (see below), and therefore the occupation
numbers will be somewhat smaller,
\begin{equation}\label{occcub}
n_k \sim O(10)\,  \lambda^{-2} \ .
\end{equation}

Because of the nonlinear dependence of the tachyonic mass on
$\phi$, a detailed description of this process is more involved
than in the theory (\ref{aB1}). Indeed, even though the typical
amplitude of the growing fluctuations is given by (\ref{typical}),
the speed of the growth of the fluctuations increases considerably
if the initial amplitude is somewhat bigger than (\ref{typical}).
Thus even though fluctuations with an amplitude a few times
greater than (\ref{typical}) are exponentially suppressed, they
grow faster and may therefore have a greater impact on the process
than fluctuations with  amplitude (\ref{typical}).

Low probability fluctuations with $ \delta \phi \gg \delta\phi_{\rm
rms}$  correspond to peaks of the initial Gaussian distribution of
the fluctuations of the field $\phi$. 
 The theory of the 3d random Gaussian fields 
is well developed   \cite{BBKS}.    Its statistical properties are
determined by the spectrum $|\delta \phi_k|^2$. One of the most
interesting features of the Gaussian field is
statistics and the shapes of the high peaks of the field distribution.
Such peaks tend to be spherically symmetric.  As a result, the whole
process looks not like a uniform growth of all modes, but more
like  bubble production (even though there are no instantons in
this model).  A simple physical interpretation of the inhomogeneous fragmentation
of the field $\phi$  is based on the fact that the interaction $-\lambda v\phi^3$ 
corresponds to  attraction  between the fluctuation modes.
As a result, the seed inhomogeneities (the peaks of the initial random
distribution) will be amplified due to the nonlinear  interaction of the fluctuations. A well known example of this type of instability
is gravitational instability of matter in the universe.

To study the growth of fluctuations in a more detailed way, one may use the stochastic
approach to tunneling and bubble formation developed in \cite{Linde2}. The
main idea of this approach can be explained as follows. Tunneling can be
represented as a result of the accumulation of quantum fluctuations whose
amplitude greatly exceeds their usual value determined by the uncertainty
principle. This happens when the long wavelength  quantum fluctuations
responsible for the tunneling correspond to bosonic excitations with large
occupation numbers. In such cases one can treat these fluctuations as
classical fields  experiencing Brownian motion due to their interaction
with the short wavelength quantum fluctuations.

Suppose that the large fluctuations of the scalar field responsible for
reheating in the model (\ref{cub}) initially look like spherically
symmetric bubbles (which is the case if the probability of such
fluctuations is strongly suppressed, see above). The equation of motion for a
bubble of a scalar field $\phi(r)$ in  Minkowski space is
\begin{equation}\label{10}
\ddot\phi = \phi''  +  2  \phi' r^{-1}  - V'(\phi) \  .
\end{equation}
Here $r$ is a distance from the center of the bubble and $\phi' =
{\partial\phi\over \partial r}$.
At the moment of its formation, the bubble wall does not move, $\dot\phi =
0$, $\ddot\phi = 0$ (critical bubble). Then it gradually starts
growing,  $\ddot\phi > 0$, which requires that
\begin{equation}\label{11}
 | \phi'' +  2 \phi' r^{-1}| < - V'(\phi) \ .
\end{equation}
A bubble of a classical field is formed only if  it
contains a sufficiently large field $\phi$, and if the bubble itself is
sufficiently large. If the size of the
bubble is too small, the gradient terms are greater than the term
$|V'(\phi)|$, and
the field $\phi$ inside the bubble does not grow.

At small $r$ the shape of the bubble can be approximated by $\phi = \phi(0)
-\alpha r^2/2$. In this approximation, the bubble has a typical size $r_0
\sim \sqrt{2\phi(0)\over \alpha}$, and  $\phi' r^{-1} = \phi'' = -\alpha $.
Therefore at the moment of the bubble formation, when $\ddot \phi = 0$, one
has
\begin{equation}\label{11a}
 \phi''   = V'(\phi(0))/3 \ .
\end{equation}
Replacing $\phi''$ by $k_0^2 \phi(0)$ one finds that the bubble can be
considered a result of overlapping of quantum fluctuations with typical
momenta $k \lesssim k_0 \sim r_0^{-1}$, where
\begin{equation}\label{11aa}
 k^{2}_0 = C^2 {V'(\phi(0))\over 3\phi(0) } \ .
\end{equation}
Here $C = O(1)$ is some numerical factor reflecting uncertainty in our
estimate of $k_0$.

Let us estimate the probability of an event when vacuum fluctuations
occasionally build up a configuration of the field satisfying this
condition.
In order to do it one should remember that the dispersion of quantum
fluctuations of the  field $\phi$ with $k < k_0$ is given by $\langle
\delta\phi^2 \rangle \sim {k_0^2\over 8\pi^2}$.
This gives
\begin{equation}\label{14}
\langle \delta\phi^2 \rangle_{k<k_0} \sim {k_0^2\over 8\pi^2} =  C^2
{V'(\phi(0))\over 24 \pi^2 \phi(0) } \ .
\end{equation}

This is an estimate of the dispersion of perturbations that may sum up
to produce a bubble of the field $\phi$ that satisfies the condition
(\ref{11}).  Of course,
this estimate is rather crude.  But
let us  nevertheless use eq. (\ref{14}) to evaluate the probability that
these
fluctuations build up a bubble of a radius  $r \gtrsim
k^{-1}_0$ containing the field $\phi$ at its center. Assuming, in the first
approximation, that the probability distribution is gaussian, one finds:
\begin{equation}\label{15}
P(\phi) \sim \exp\left(-{\phi^2\over
2\langle \delta\phi^2 \rangle_{k<k_0}}\right) =
\exp \left(-{12\pi^2 \phi^3\over C^2 V'(\phi)}\right) \ .
\end{equation}

Let us first apply this result to the theory $-\lambda\phi^4/4$. In this
case one finds
\begin{equation}\label{15a}
P(\phi) \sim
\exp \left(-{12\pi^2 \over C^2 \lambda }\right) \ .
\end{equation}
Note that the factor in the exponent in (\ref{15a}) to within a factor of
$C \approx 2$  coincides with the Euclidean action $S_E$ in eq. (\ref{4}).
Taking into account  the very rough method we used to estimate $k_0$ and
calculate the dispersion of the perturbations responsible for tunneling,
the coincidence is rather impressive. It was shown in
\cite{Star,Linde2} that this approach gives exactly the same
answer as the Euclidean approach for the case of tunneling during inflation
when $V'' \ll H^2$.

Most importantly, this method allows us  to investigate tunneling and
the development of instability in
the theories where instanton solutions do not
exist \cite{Linde2}. In particular, for tunneling in the theory
$-\lambda v \phi^3/3$ one finds
\begin{equation}\label{16}
P(\phi) \sim  (\lambda v \phi)^2  \exp \left(-{12\pi^2 \phi \over C^2 \lambda
v}\right) \ .
\end{equation}
We included here the subexponential factor $O(k_0^4) \sim (\lambda v
\phi)^2$, which is necessary to describe the probability of tunneling per
unit time per unit volume.

This means that tunneling is not suppressed for $\phi \sim {C^2 \lambda
v\over 12\pi^2}$. This result is in agreement with our previous estimate
(\ref{typical}). Now let us take into account that the total time of the
development of instability is a sum of the time of tunneling plus the time
necessary for rolling of the field down. One can show that the time of
rolling down is inversely proportional to $m(\phi) \sim \sqrt{\lambda v
\phi}$, i.e. it decreases at large $\phi$. Also, the subexponential factor
$(\lambda v \phi)^2$ grows at large $\phi$, which makes tunneling to large
$\phi$ faster.  Consequently, as we already discussed above, the main
contribution to the development of instability is given by the fluctuations
with $\phi  \gtrsim {C^2 \lambda v\over 12\pi^2}$. Exponential suppression
of the probability of such fluctuations leads to their approximate
spherical symmetry.

The results of our lattice simulations for this model
are shown in  Fig. \ref{2gif}.  In this model bubbles form quickly
enough (unlike in the model $-\lambda \phi^4$), so  we were able to start with quantum fluctuations centered at
$\langle\phi\rangle =0$ and allow the bubbles to form. The bubbles (high
peaks of the
field distribution) grow, change shape, and interact with each
other, rapidly  dissipating the vacuum energy $V(0)$.
\newpage

\begin{figure}[Fig001]
\leavevmode\epsfysize=6.5cm \epsfbox{\picdir{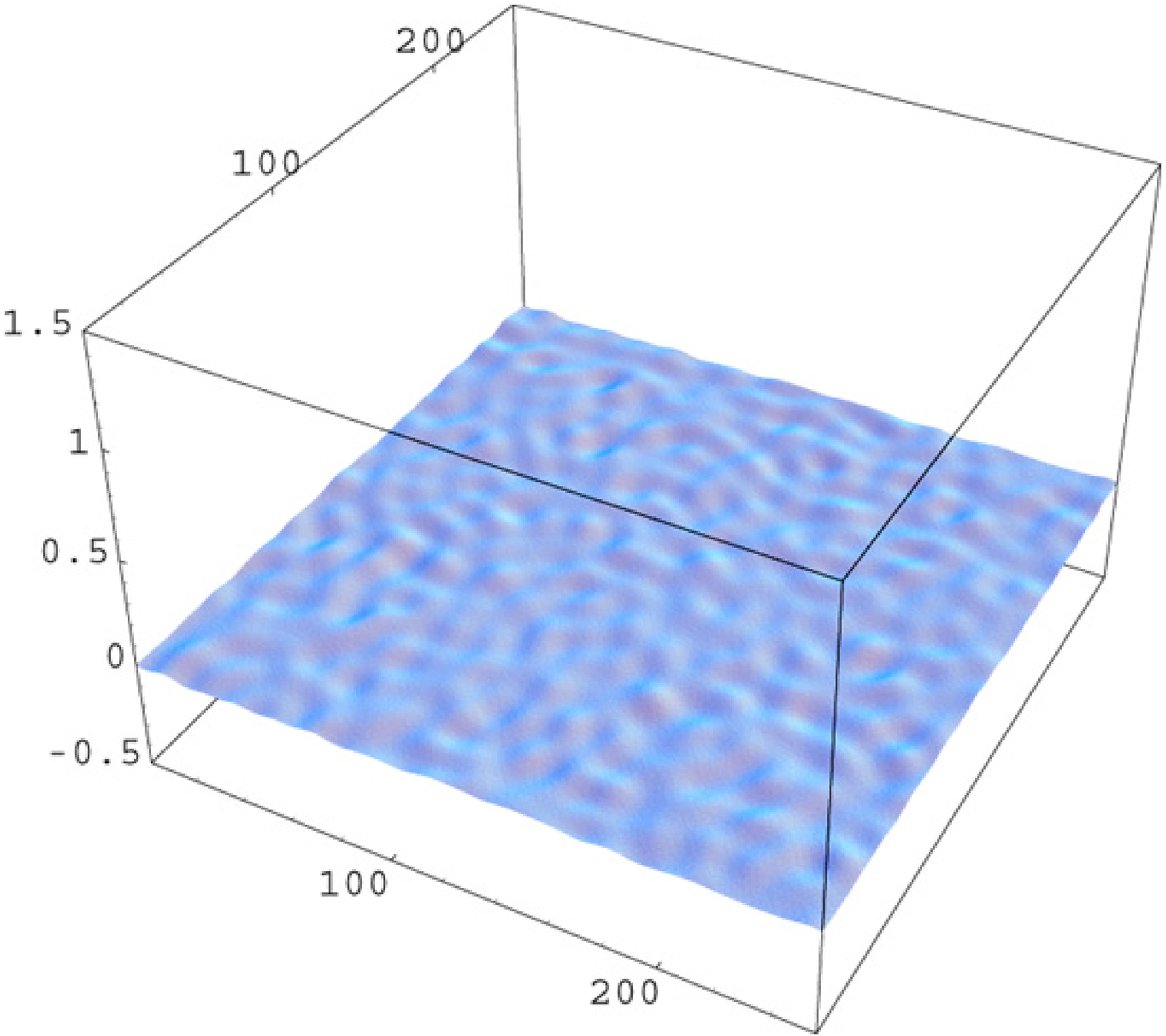}  }\hskip 1 cm
\leavevmode\epsfysize=6.5cm  \epsfbox{\picdir{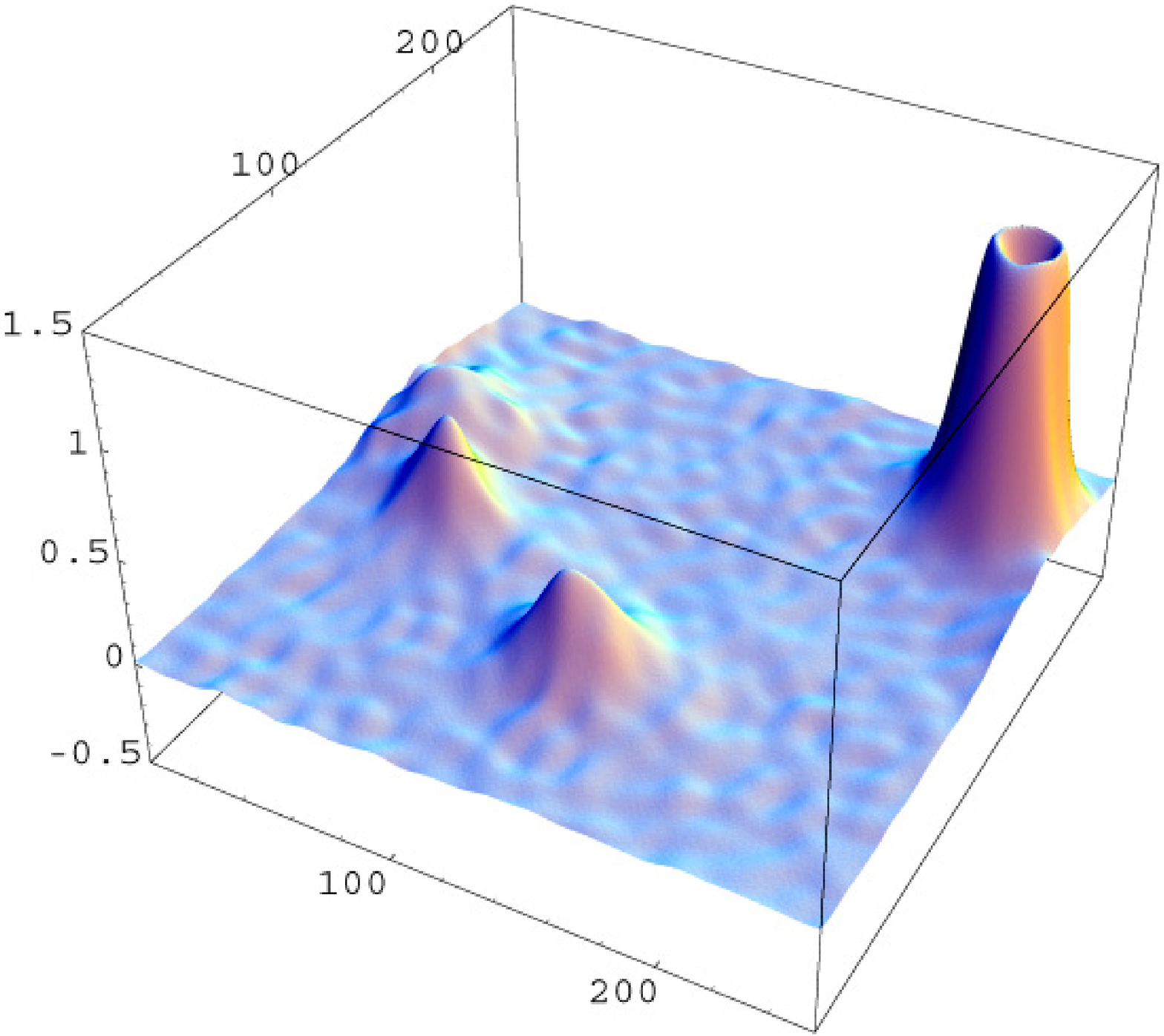}}
\end{figure}
\vskip -0.8cm

\begin{figure}[Fig001]
\leavevmode\epsfysize=6.5cm \epsfbox{\picdir{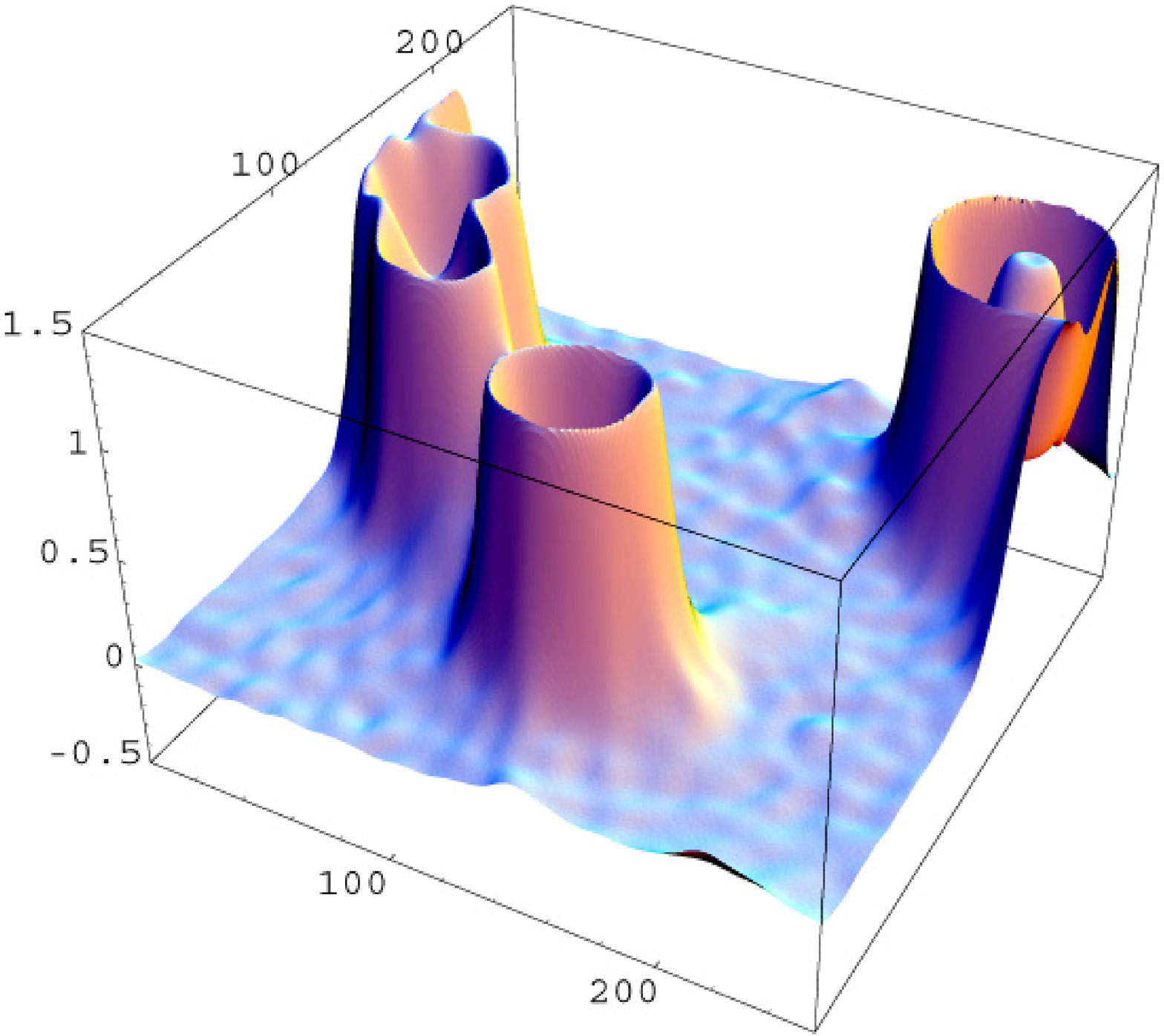}  }\hskip 1 cm
\leavevmode\epsfysize=6.5cm  \epsfbox{\picdir{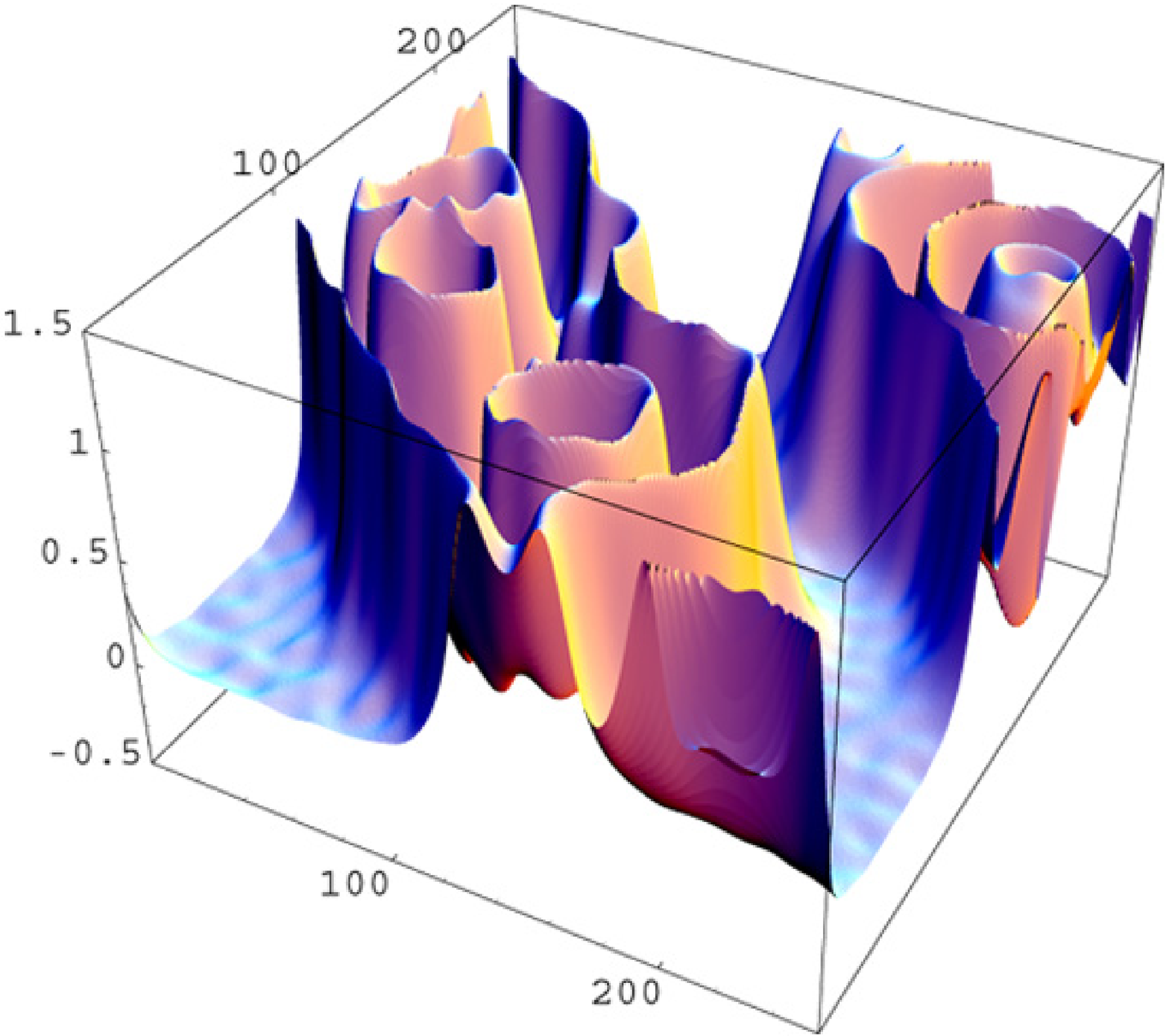}}
\end{figure}
\vskip -0.8cm
\begin{figure}[Fig001]
\leavevmode\epsfysize=6.5cm \epsfbox{\picdir{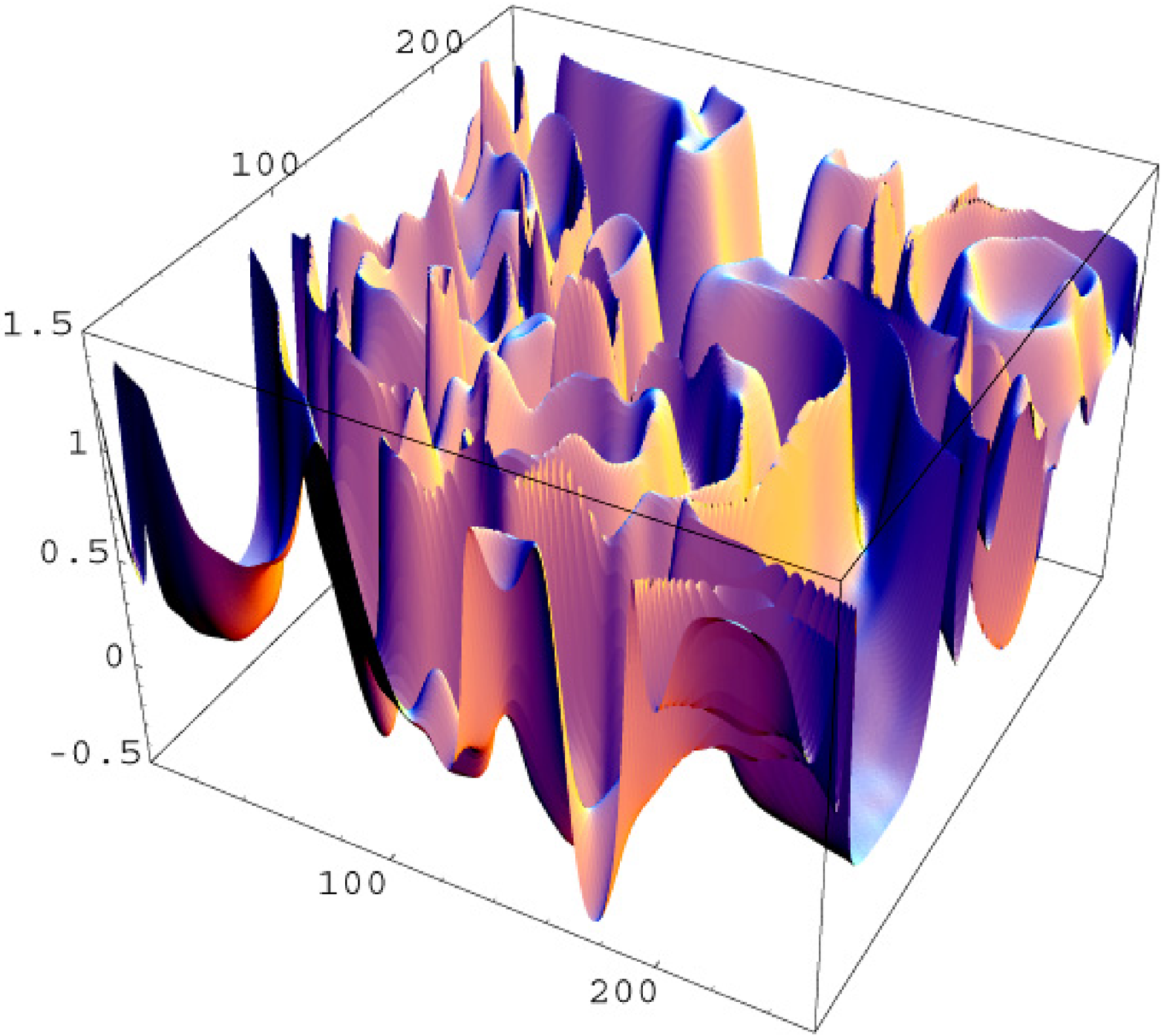}  }\hskip 1 cm
\leavevmode\epsfysize=6.5cm  \epsfbox{\picdir{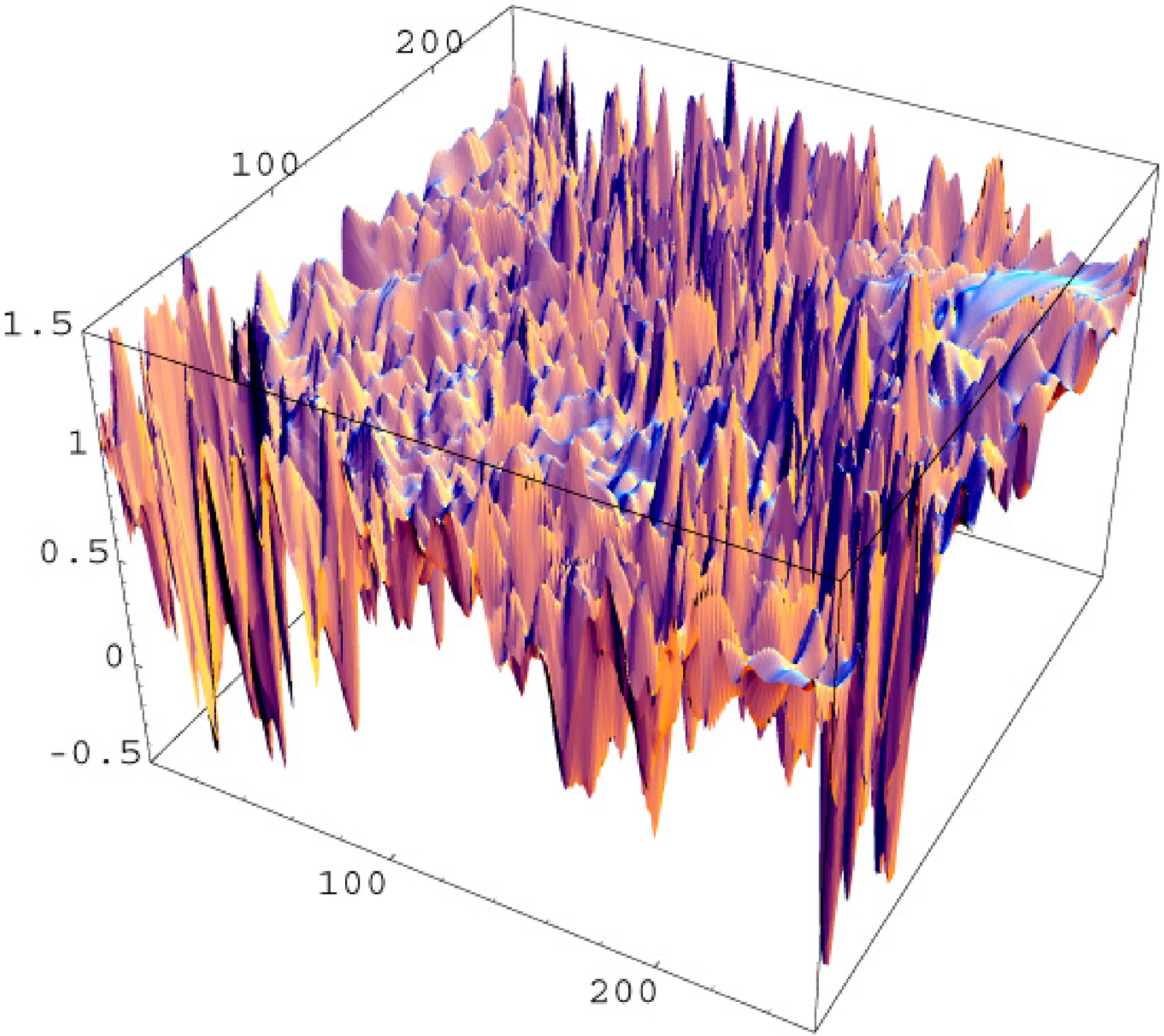}}

\

\caption[Fig001]{\label{2gif} Field values on a 2D slice through the lattice for $V= -{\lambda\over 3} v\phi^3+{\lambda\over 4}\phi^4$ (\ref{cub}). The growth of quantum fluctuations of $\phi$
looks like bubble formation. Remarkably, the bubbles expand and collide even before the average field value reaches the minimum. Preheating occurs due to a
combined effect of bubble production, tachyonic instability and bubble wall
collisions. This figure should be compared with Fig. \ref{onefieldslice} for the
theory $V= -{m^2\over 2}\phi^2+{\lambda\over 4}\phi^4$ (\ref{aB1}). }
\end{figure}

\newpage

\begin{figure}[Fig002]
\centering \leavevmode\epsfysize=12cm \epsfbox{\picdir{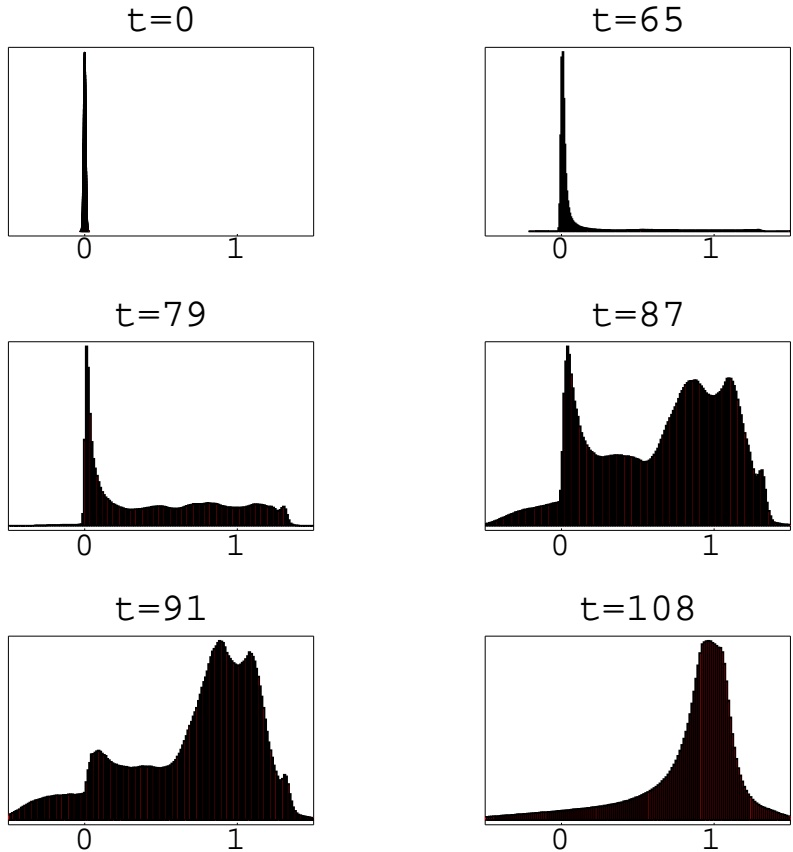}}

\

\caption[Fig002]{\label{cubicdistrib} Histograms describing the process of
symmetry breaking in the model (\ref{cub}) for $\lambda = 10^{-2}$. After reaching the minimum of the effective potential, the distribution acquires the form shown in the last  frame and  practically does not oscillate.  The last histogram corresponds to the last frame in Fig. \ref{2gif}.}
\end{figure}

Figure \ref{cubicdistrib} shows the probability distribution
$P(\phi,t)$ in the model (\ref{cub}). As we see, in this model the
field distribution also rapidly relaxes near the minimum of the effective potential within a single oscillation. In this case the histogram in the beginning looks pretty chaotic because of bubble formation and bubble wall collisions, which we could see in the previous figure.

One should note that the numerical investigation of this model
involved specific complications due to the necessity of performing
renormalization. Lattice simulations involve the study of modes
with large  momenta that are limited by the inverse lattice
spacing. These modes give an additional contribution to the
effective parameters of the model.   In the limit of zero lattice
spacing these corrections would become infinite, but they are regularized
by the lattice cutoff.  In our simulations of the simple model (\ref{aB1})
these corrections gave a contribution to the effective mass of the field
$\phi$ that was much smaller than $m$ and therefore did not affect our
results. Meanwhile, in the cubic model similar corrections
 induce a (fictitious) linear term $\lambda v \phi
\langle\phi^2\rangle$. This term should be subtracted by the
proper renormalization procedure, which brings the effective
potential back to its form (\ref{cub}). See the appendix for more details. This was the first time in our simulations when a careful treatment of
high frequency modes was necessary.  A similar situation  may occur in any
theory where $V''(0) = 0$, such as the theory $-\lambda\phi^4$ discussed in
the previous section.

\begin{figure}[Fig001]
\centering \leavevmode\epsfysize=8cm
\epsfbox{\picdir{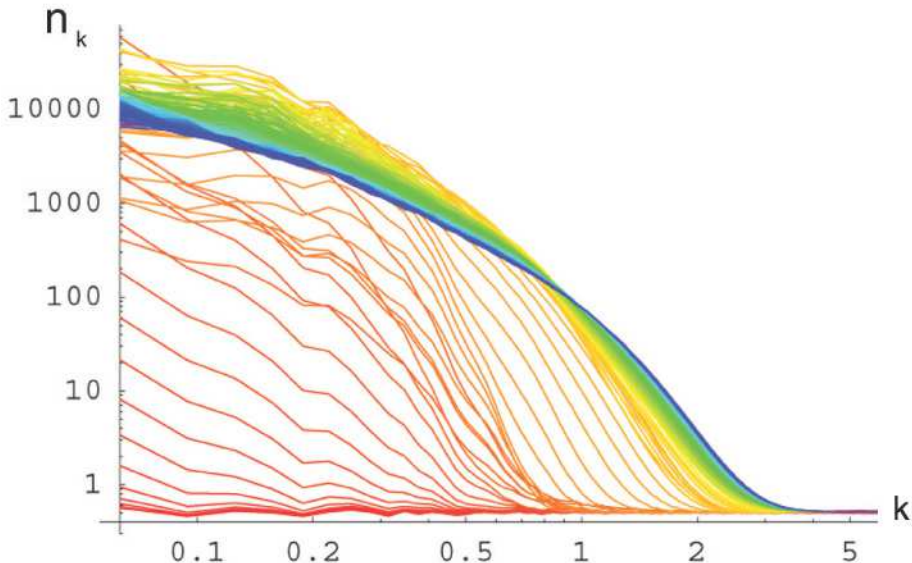}}

\

\caption[Fig001]{\label{cubicocc} {Occupation numbers of particles produced
during tachyonic preheating in the model (\ref{cub}) with $\lambda =
10^{-2}$.}}
\end{figure}

Figure \ref{cubicocc} shows the occupation numbers of produced particles in
the model (\ref{cub}) with $\lambda = 10^{-2}$. These occupation
numbers grow up to $10^4$ -- $10^5$ within a single oscillation, which is
in good agreement with our estimate (\ref{occcub}).

\section{Conclusions}

In this paper we studied the dynamics of spontaneous symmetry breaking, which
occurs when a scalar field falls down from the top of its effective
potential. We have found, in agreement with \cite{GBFKLT}, that   the main
part of this process typically completes within a single oscillation  of
the distribution of the scalar field. This is a very unexpected conclusion
that may have important cosmological implications.

One of the most efficient mechanisms for the creation of matter after inflation
in theories with convex effective potentials ($V''(\phi) > 0$) is the
mechanism of parametric amplification of vacuum fluctuations in the process
of homogeneous oscillations of the inflaton field, which was called
preheating \cite{KLS}. It has also been noted that in the case where
potentials become concave ($V''(\phi) < 0$), preheating may become more
efficient \cite{Prokopec}. Now we see that this effect is very generic. In
many theories with concave potentials the energy of an unstable vacuum
state is transferred to the energy of inhomogeneous classical waves of
scalar fields  within a single oscillation of the field distribution. We
emphasize here that we are talking about the oscillations of the field
distribution rather than about the oscillations of a homogeneous field
$\phi$ because quite often the homogeneous component $\langle \phi \rangle $ of the field $\phi$
remains zero during the process of spontaneous symmetry breaking.

One of the important consequences of our results is the observation \cite{GBFKLT} that in many models of hybrid inflation  \cite{Hybrid} the first stage of  reheating occurs not due to  homogeneous oscillations of the scalar field  but due to tachyonic preheating \cite{GBFKLT}. A detailed discussion of this effect will be contained in \cite{GBFKL}.

The process of preheating and symmetry breaking may take an especially unusual form in the theory of brane inflation \cite{Dvali:2001fw,Burgess:2001fx} based on the hybrid inflation scenario and the mechanism of tachyon condensation on the brane antibrane system \cite{Sen}.

The situation in models of the type used in the new inflation scenario
is somewhat more complicated. In these models the potential is also
concave. However, the expansion of the universe stretches inhomogeneities of
the field rolling down from the top of the effective potential and makes it
homogeneous on an exponentially large scale. Therefore to evaluate a
possible significance of tachyonic instability in this regime one must
compare the amplitude of the homogeneous component of the field with the
amplitude of the quantum fluctuations.  The result appears to be  very
sensitive to the scale  of spontaneous symmetry breaking in such models. A
preliminary  investigation of this issue indicates that in small-field
models where  the scale of spontaneous symmetry breaking  is much smaller
than $M_p$, the leading mechanism of preheating typically is tachyonic. If
correct, this would be a very interesting conclusion  indicating
that in large-field models the leading mechanism of preheating
typically is related to parametric resonance, whereas in small-field models  the main mechanism of preheating is typically tachyonic, at least at the first stages of the process.  We will return to
the discussion of this issue in a coming publication  \cite{FKL2}.

Finally we should mention that an interesting application of our methods can be found in the recently proposed  ekpyrotic and pyrotechnic scenario \cite{KOST,KKL}. Even though we are very skeptical with respect to the ekpyrotic/pyrotechnic scenario for many reasons explained in \cite{KKL}, it is still interesting that the methods developed in the theory of tachyonic preheating provide us with a very simple theory of the generation of density perturbations in these models \cite{KKL}.

It is a pleasure to express our gratitude to J. Garc\'\i a-Bellido, P.~B.~Greene,   and I.~Tkachev for the collaboration at the early stages of this project. We thank NATO Linkage Grant 975389 for support.  L.K. was supported
by NSERC and by CIAR; G.F. and A.L. were supported
by NSF grant PHY-9870115 and by the Templeton Foundation.

\appendix
\section{The Lattice Calculations}

\subsection{Overview}

The lattice calculations reported on in this article were all done using the program LATTICEEASY, developed by Gary Felder and Igor Tkachev. The program records the value of the fields and derivatives at each point on a spatial grid with evenly spaced points. The fields are then evolved using their classical equations of motion 
\begin{equation}
\ddot \phi -\nabla_{\vec x}^2  \phi+V'=0.
\end{equation}
 The use of the classical equations is justified because the instability discussed rapidly drives the fields to a state with exponentially large occupation numbers, meaning they effectively act as classical fields  \cite{lattice}. Although LATTICEEASY is designed to (optionally) include the effects of cosmological expansion on field evolution all of the simulations reported here were done in a  flat  spacetime  background. The effects of expansion will be discussed in subsequent publications (\cite{FKL2,GBFKL}).

Time evolution is done with a staggered leapfrog algorithm using a fixed time step. The initial conditions for the fields and derivatives are set in momentum space and then Fourier transformed to give the initial spatial distribution. The initial values of the modes are given by quantum fluctuations. Each mode has a random phase and a Gaussian random amplitude with expectation value
\begin{equation}
\langle\vert\phi_k\vert^2\rangle = {1 \over 2 k}.
\end{equation}
The exception to this is the Coleman-Weinberg potential for which the initial conditions were set by the instanton configuration described in the text. Note that ordinarily the equations for quantum fluctuations would have $\sqrt{k^2 + m^2}$ in the place of $k$ in the formula above. For the models discussed here, however, we were simulating a quench in which the effective squared mass of the fields is presumed to have rapidly become negative. Thus we used initial conditions corresponding to massless fluctuations. For some of the runs here we imposed a momentum cutoff, setting $\phi_k=0$ for all modes above a certain momentum $k$. Such cutoffs eliminated unphysical effects from quantum fluctuations that were not excited to large values. In each such case we also ran without the cutoff and found the results to be qualitatively similar except for the addition of high frequency noise in the field distribution.

The plots shown in this paper show either field values (which are self-explanatory), probability distribution functions (PDF), or occupation number spectra. The PDF of a field is obtained by dividing the field values on the grid into evenly spaced bins and simply counting the number of gridpoints in which the field value was in each bin. The occupation number is defined by Fourier transforming the field and computing for each mode  by Eq. (\ref{number}),
where $m^2 \equiv  V''$,
$\omega_k \equiv \sqrt{k^2 + m^2}$ for $m^2> 0$. For $m^2 < 0$ one can use either   $\omega_k \equiv |k|$ or $\omega_k \equiv \sqrt{k^2 + |m^2|}$. 
All plots shown in this paper except Fig. 1 and Fig. 2 use $\omega_k = \vert k\vert$ when $m^2 < 0$ but the results were qualitatively similar using either definition. The occupation number $n_k$ is given by averaging over a spherical shell in Fourier space.
 This definition coincides with the standard one in the end of the process, where the mass squared becomes positive and topological defects disappear.

The full details of these lattice calculations can be found in the documentation available on the LATTICEEASY website at
~http://physics.stanford.edu/latticeeasy. Moreover, these calculations have been discussed in previous publications of ours (e.g. \cite{lattice,FT}). This is our first publication where we discuss simulations that used renormalization, however, so we will discuss this procedure in the next section. The last section of the appendix lists the parameters used for each of the runs illustrated in the paper.

\subsection{Renormalization}

As we have discussed, the justification for doing a classical
calculation for quantum fields is that once the field fluctuations
are amplified sufficiently quantum effects are negligible. There
are some cases, however, when these quantum effects may be
important, and in such cases they may be (partially) accounted for
through a simple form of renormalization.

Consider how this applies to the lattice calculations discussed
here. 
Initially the field fluctuations are only those representing
quantum vacuum states. These fluctuations affect couplings,
masses, and the total energy of the system in a way that is
dependent on the lattice spacing. For example, consider the theory
\begin{equation}
V = {1 \over 4} \lambda \phi^4 - {1 \over 3} \lambda v \phi^3 + {1 \over 12} \lambda v^4
\end{equation}
and rewrite the field $\phi(x,t)$ as the sum of a homogeneous component $\phi(t)$ and fluctuations $\delta\phi$. The effective potential felt by the homogeneous field $\phi$ will receive a correction (among others) from the fluctuations equal to
\begin{equation}
\delta V \approx -\lambda v \langle\delta\phi^2\rangle \phi.
\end{equation}
(The $1/3$ is cancelled by a coefficient arising from combinatorics.) This correction represents an unphysical effect in the sense that its strength depends on the ultraviolet cutoff imposed by the lattice. In the limit of zero lattice spacing where arbitrarily large momenta would be included on the lattice this correction would become infinite. This would add an unphysical term $C\phi$ to the effective potential. This   effect can be eliminated, however, by adding a counterterm
\begin{equation}
\Delta V  = \lambda v \langle\delta\phi^2\rangle \phi
\end{equation}
or equivalently by adding the term
\begin{equation}
\lambda v \langle\delta\phi^2\rangle
\end{equation}
to the equation of motion for $\phi$. Note that $\langle\delta\phi^2\rangle$ in this case refers to the value that arises from initial quantum fluctuations, not to a dynamic quantity that changes as the field evolves and fluctuations grow. Such changes represent physical effects and should not be eliminated. In effect this correction eliminates the linear term in the potential at $\phi=0$ when the field is in the vacuum state.

The above example illustrates how a simple form of renormalization can be implemented on the lattice. This procedure could in principle be used to renormalize any mass, coupling constant, or energy term in the theory. Ordinarily these corrections are not important (unless one uses a very large value of the momentum cutoff) because the quantum effects are quickly swamped as the fluctuations become amplified. We did not find it necessary to use renormalization for any models except the cubic one, where we used it as described here to prevent the field from artificially rolling away from $\phi=0$ due to the induced linear term.

\subsection{List of Parameters}

In this section we list the parameters used for the lattice simulations from which all of the figures in the paper were drawn. The models discussed in the paper will be referred to here simply as Quadratic (model \ref{aB1}), Complex (model \ref{aB1} with $\phi^2$ replaced by $\vert\phi\vert^2$), Quartic (model \ref{CW}), and Cubic (model \ref{cub}). The parameters $N$, $L$, $dt$, and $k_{cut}$ refer to the number of gridpoints, the width of the box, the time step, and the initial momentum cutoff respectively. Lengths and times are measured in units of $\sqrt{\lambda} v$. The coupling constant $\lambda$ is also given for each run, as well as any information specific to the particular plot. All runs are assumed to be three dimensional unless otherwise indicated.

\begin{description}

\item[Figures \ref{onefielddistrib}, \ref{onefieldslice}, and \ref{onefielddistribdispl}] Quadratic model, $N = 256^3$, $L = 100$,
$dt = ~.1$, $k_{cut} = 1$, $\lambda = 10^{-4}$. The run shown in Fig. \ref{onefielddistribdispl} started with a homogeneous displacement $\langle\phi\rangle=.1 v$. The PDF's in figure \ref{onefielddistrib} and \ref{onefielddistribdispl} use $256$ bins.

\item[Figure \ref{domains}] Quadratic model, Two dimensional simulation with $N = 1024^2$, $L = 1600$, $dt =~.1$, $k_{cut} = 0$, $\lambda = 10^{-2}$. The slices shown are plots of $128^2$ points where each point represents an average over an $8\times 8$ box of gridpoints. This averaging was done simply to make plotting easier and reduce the resulting image sizes.

\item[Figures \ref{complex} and \ref{compocc}] Complex model, $N = 256^3$, $L = 600$, $dt =~.2$, $\lambda = 10^{-4}$. The momentum cutoff was $k_{cut}=2$ and $k_{cut}=0$ for Figs. \ref{complex} and \ref{compocc} respectively. The PDF's in Fig. \ref{complex} used $80\times 80$ bins for the 2D field space. The spectra shown in Fig. \ref{compocc} are for the real component ${\rm Re}(\phi)$. The spectra for ${\rm Im}(\phi)$ look  identical.

\item[Figure \ref{strings}] Complex model, $N = 128^3$, $L = 40$, $dt = .01$, $k_{cut}=\sqrt{2}$, $\lambda = 10^{-4}$. A string was defined as the collection of points on the lattice for which $\vert\phi\vert^2 < .02$.

\item[Figures \ref{instJPEG} and \ref{instJPEG2}] Quartic model, $N = 256^3$, $L = 500$, $dt = .1$, $k_{cut}=0$, $\lambda = 5 \times 10^{-5}$. Both runs used as initial conditions the instanton solution (\ref{3aa}) at $t=0$ with $\phi_0=.02 v$. The run shown in Fig. \ref{instJPEG} did not include additional quantum fluctuations, while the run shown in Fig. \ref{instJPEG2} did. Both figures show a partial 2D slice through the lattice, cut so as to show the interior of the bubble.

\item[Figures \ref{2gif}, \ref{cubicdistrib}, and \ref{cubicocc}] Cubic
model, $N = 256^3$, $L = 200$, $dt = ~.25$, $\lambda = 10^{-2}$. The
momentum cutoff $k_{cut}$ was $.6$ for Figs. \ref{2gif} and
\ref{cubicdistrib} and $0$ for figure \ref{cubicocc}. The PDF*s in Fig. \ref{cubicdistrib} used $256$ bins.

\end{description}

\end{document}